\documentstyle[12pt]{article}
\makeatletter
\@addtoreset{equation}{section}
\makeatother

\newlength{\extraspace}
\newlength{\extraspaces}
\newcommand{\be}{\begin{equation}
\addtolength{\abovedisplayskip}{\extraspaces}
\addtolength{\belowdisplayskip}{\extraspaces}
\addtolength{\abovedisplayshortskip}{\extraspace}
\addtolength{\belowdisplayshortskip}{\extraspace}}
\newcommand{\ee}{\end{equation}}
\newcommand{\ba}{\begin{eqnarray}
\addtolength{\abovedisplayskip}{\extraspaces}
\addtolength{\belowdisplayskip}{\extraspaces}
\addtolength{\abovedisplayshortskip}{\extraspace}
\addtolength{\belowdisplayshortskip}{\extraspace}}
\newcommand{\ea}{\end{eqnarray}}
\newcommand{\nonu}{\nonumber \\[.5mm]}
\newcommand{\A}{&\!\!\!}

\begin{document}
\thispagestyle{empty}
\setlength{\baselineskip}{6mm}
%
\vspace*{5mm}
\begin{center}
{\large\bf Linearizing extended nonlinear supersymmetry \\[2mm]
in two dimensional spacetime 
} \\[20mm]
{\sc Motomu Tsuda}
\footnote{
\tt e-mail: motomu.tsuda@gmail.com} 
\\[5mm]
{\it Aizu Hokurei High School \\
Aizuwakamatsu, Fukushima 965-0031, Japan} \\[20mm]
\begin{abstract}
We linearize nonlinear supersymmetry in the Volkov-Akulov (VA) theory for extended SUSY 
in two dimensional spacetime ($d = 2$) based on the commutator algebra. 
Linear SUSY transformations of basic component fields for general vector supermultiplets are uniquely determined 
from variations of functionals (composites) of Nambu-Goldstone (NG) fermions, 
which are represented as simple products of powers of the NG fermions and a fundamental determinant in the VA theory. 
The structure of basic component fields with general auxiliary fields in the vector supermultiplets 
and transitions to $U(1)$ gauge supermultiplets through recombinations of the functionals of the NG fermions 
are explicitly shown both in $N = 2$ and $N = 3$ theories as the simplest and general examples for extended SUSY theories in $d = 2$. 
\\[5mm]
%
%
\noindent
PACS:11.30.Pb, 12.60.Jv, 12.60.Rc \\[2mm]
\noindent
Keywords: supersymmetry, Nambu-Goldstone fermion, commutator algebra, vector supermultiplet 
\end{abstract}
\end{center}

\newpage

\section{Introduction}

Relations between Volkov-Akulov (VA) nonlinear supersymmetric (NLSUSY) theory \cite{VA} and linear SUSY (LSUSY) ones \cite{WZ,WB} 
are shown explicitly for $N = 1$ and $N = 2$ SUSY theories \cite{IK}-\cite{STT2}. 
In superspace formalism, superfields on specific superspace coordinates which depend on Nambu-Goldstone (NG) fermions \cite{IK} 
give systematically the general relation between the VA NLSUSY and LSUSY theories (NL/LSUSY relation) 
as demonstrated in $N = 1$ SUSY theories \cite{IK}-\cite{STT1}. 
In the component expression, it is possible to construct heuristically functionals (composites) of the NG fermions, 
which reproduce LSUSY transformations of basic component fields under their NLSUSY transformations, 
and its heuristic method was used to study the NL/LSUSY relation for $N = 2$ minimal $U(1)$ gauge supermultiplet \cite{STT2} 
(and also for a $N = 3$ LSUSY theory in two dimesional spacetime ($d = 2$) \cite{ST0}), 
though the constructions of the functionals in all orders of the NG fermions are complicated problems. 

On the other hand, we have recently proposed a linearization procedure of NLSUSY 
based on a commutator algebra in the VA NLSUSY theory \cite{MT1} by introducing a set of fermionic and bosonic functionals 
which are represented as products of powers of the NG fermions and a fundamental determinant 
indicating a spontaneous SUSY breaking in the VA NLSUSY action. 
In this linearization method, variations of basic components defined from the above set of the functionals 
under the NLSUSY transformations uniquely determine LSUSY transformations for (massless) vector supermultiplets.  
This is based on a fact that every functional of the NG fermions and their derivative terms 
satisfies the same commutation relation in the VA NLSUSY theory. 
Moreover, we have shown in $N = 1$ SUSY theories that $U(1)$ gauge and scalar supermultiplets 
in addition to a vector one with general auxiliary fields are derived from the same set of the functionals 
and their appropriate recombinations \cite{MT2}. 

Because the all-order functional (composite) structure of the NG fermions for vector supermultiplets is manifest 
in the commutator-based linearization of NLSUSY, 
its procedure would be useful to understand the NL/LSUSY relation for $N \ge 2$ extended SUSY in more detail 
and furthermore to know low-energy physics of a NLSUSY general relativisitic (GR) theory \cite{KS}. 
The Einstein-Hilbert-type (global) NLSUSY fundamental action in the NLSUSY-GR theory possesses rich symmetries \cite{ST1}, 
which are isomorphic to $SO(N)$ super-Poincar\'e group and contains the VA NLSUSY action in the cosmological term. 
Therefore, it is important for the $N$-extended NLSUSY-GR theory and its composite model interpretation 
({\it {superon-quintet model (SQM)}}) \cite{KS0} in the low energy 
to study more extensively and explicitly the NL/LSUSY relations for $N$-extended SUSY. 

In this paper, we focus on the VA NLSUSY theory for extended SUSY in $d = 2$ 
and apply the commutator-based linearization procedure to it 
since explicit calculations for the $d = 2$ NL/LSUSY relation give 
many suggestive and significant results to $d = 4$ SUSY theories 
as for the structure of the functionals of the NG fermions for vector supermultiplets \cite{ST2} 
and SUSY models with interaction terms \cite{ST3}-\cite{ST5} etc. 
In Section 2, as in the case of the linearization in $d = 4$ \cite{MT1}, 
we introduce a set of bosonic and fermionic functionals for vector supermultiplets in the $d = 2$ VA NLSUSY theory, 
which are represented as products of powers of the NG fermions and the fundamental determinant, 
and explain some properties of those functionals. 
In Section 3, we show LSUSY transformations with general auxiliary-field structure for vector supermultiplets, 
which are uniquely determined from basic components defined from the set of the functionals of the NG fermions 
by evaluating their variations under NLSUSY transformations based on the commutator algebra. 

In the remaining section of this paper, we construct $N = 2$ and $N = 3$ LSUSY multiplets 
from the LSUSY transformations shown in Section 3 
as typical and comprehensive examples for the extended SUSY theories. 
In Section 4, we derive a $d = 2$, $N = 2$ vector supermultiplet \cite{ST2} 
by means of the reduction from those general basic components and LSUSY transformations. 
We also discuss in Section 5 the construction of $N = 2$ minimal $U(1)$ gauge and scalar supermultiplets 
by using appropriate recombinations of the functionals of the NG fermions in the $N = 2$ vector supermultiplet. 
Those arguments in the linearization of $N = 2$ NLSUSY give us valuable lessons for $N \ge 3$ SUSY theories. 

In Section 6, as the simplest but a general extended LSUSY model in $d = 2$, 
we study the structure of the basic components with general auxiliary fields in a $N = 3$ vector supermultiplet 
by focusing procedures for counting degrees of freedom (d.o.f.) of bosonic and fermionic components. 
In Section 7, we show a transition from the $N = 3$ vector supermultiplet to a minimal $U(1)$ gauge one 
by means of general recombinations of the functionals of the NG ferimions, 
which correspond to a generalization of the Wess-Zumino gauge to the NL/LSUSY relation for extended SUSY. 
Summary and discussions are given in Section 8.

\section{Functionals of Nambu-Goldstone fermions for vector supermultiplets}

A fundamental action in the VA NLSUSY theory \cite{VA} for extended SUSY in $d = 2$ is given in terms of 
Majorana NG fermions $\psi^i$ as 
\footnote{
The indices $i,j, \cdots = 1, 2, \cdots, N$ 
and Minkowski spacetime indices are denoted by $a, b, \cdots = 0, 1$. 
Gamma matrices satisfy $\{ \gamma^a, \gamma^b \} = 2 \eta^{ab}$ 
with the Minkowski spacetime metric $\eta^{ab} = {\rm diag}(+,-)$ 
and $\displaystyle{\sigma^{ab} = {i \over 2}[\gamma^a, \gamma^b]}$ is defined. 
}
\be
S_{\rm NLSUSY} = - {1 \over {2 \kappa^2}} \int d^2x \ \vert w \vert, 
\label{NLSUSYaction}
\ee
where $\kappa$ is a dimensional constant whose dimension is (mass)$^{-1}$ 
and a fundamental determinant $\vert w \vert$ is defined by means of 
\be
\vert w \vert = \det w^a{}_b = \det(\delta^a_b + t^a{}_b)
\label{detw}
\ee
with $t^a{}_b = - i \kappa^2 \bar\psi \gamma^a \partial_b \psi$. 
NLSUSY transformations of $\psi^i$ are 
\be
\delta_\zeta \psi^i = {1 \over \kappa} \zeta^i + \xi^a \partial_a \psi^i, 
\label{NLSUSY}
\ee
where $\zeta^i$ are constant (Majorana) spinor parameters and $\xi^a = i \kappa \bar\psi^j \gamma^a \zeta^j$. 
The NLSUSY transformations (\ref{NLSUSY}) satify a commutator algebra, 
\be
[\delta_{\zeta_1}, \delta_{\zeta_2}] = \delta_P(\Xi^a), 
\label{NLSUSYcomm}
\ee
with $\delta_P(\Xi^a)$ ($\Xi^a = 2 i \bar\zeta_1^i \gamma^a \zeta_2^i$) meaning a translation. 
Since the determinant (\ref{detw}) also transforms as 
\be
\delta_\zeta \vert w \vert = \partial_a (\xi^a \vert w \vert), 
\label{NLSUSYdetw}
\ee
the NLSUSY action (\ref{NLSUSYaction}) is invariant under NLSUSY transformations (\ref{NLSUSY}) except for total derivative terms. 

As a property of the commutator algebra (\ref{NLSUSYcomm}), 
every bosonic or fermionic Lorentz-tensor (or scalar) functional of $\psi^i$ 
and their derivative terms ($\partial \psi^i$, $\partial^2 \psi^i$, $\cdots$, $\partial^n \psi^i$) 
satisfies the commutator algebra (\ref{NLSUSYcomm}) under the NLSUSY transformations (\ref{NLSUSY}); namely, 
\be
[\delta_{\zeta_1}, \delta_{\zeta_2}] F^I_A = \Xi^a \partial_a F^I_A, 
\label{NLSUSYcomm2}
\ee
where $F^I_A = F^I_A(\psi^i, \partial \psi^i, \partial^2 \psi^i, \cdots, \partial^n \psi^i)$ are the functionals of the NG fermions 
with Lorentz index $A = (a, ab, \cdots, {\rm etc.})$ and the internal one $I = (i, ij, \cdots, {\rm etc.})$. 
\footnote{
The commutation relation (\ref{NLSUSYcomm2}) is proved from Eq.(\ref{NLSUSYcomm}) and from the fact that 
the derivative terms ($\partial \psi^i$, $\partial^2 \psi^i$, $\cdots$, $\partial^n \psi^i$) 
and products of two kinds of the fuctionals $F^I_A$ and $G^J_B$ which are respectively defined in Eq.(\ref{NLSUSYcomm2}) 
satisfy the same commutation relation (for example, see \cite{STal}). 
}

From the NLSUSY transformations (\ref{NLSUSY}) (and (\ref{NLSUSYdetw})), 
a set of bosonic and fermionic functionals of the NG fermions, whose variation under the NLSUSY transformations 
shows linear exchanges among those functionals, can be constructed by means of products of powers of $\psi^i$ 
(with $\gamma$-matrices) and the determinant $\vert w \vert$. 
Let us express the set of bosonic and fermionic functionals for each $N$ NLSUSY as 
\ba
b^i{}_A{}^{jk}{}_B{}^{l \cdots m}{}_C{}^n \left( (\psi^i)^{2(n-1)} \vert w \vert \right) 
\A = \A \kappa^{2n-3} \bar\psi^i \gamma_A \psi^j \bar\psi^k \gamma_B \psi^l \cdots \bar\psi^m \gamma_C \psi^n \vert w \vert, 
\label{bosonic} 
\\[1mm]
f^{ij}{}_A{}^{kl}{}_B{}^{m \cdots n}{}_C{}^p \left( (\psi^i)^{2n-1} \vert w \vert \right) 
\A = \A \kappa^{2(n-1)} \psi^i \bar\psi^j \gamma_A \psi^k \bar\psi^l \gamma_B \psi^m \cdots \bar\psi^n \gamma_C \psi^p \vert w \vert, 
\label{fermionic}
\ea
which mean 
\ba
\A \A 
b = \kappa^{-1} \vert w \vert, \ b^i{}_A{}^j = \kappa \bar\psi^i \gamma_A \psi^j \vert w \vert, 
\ b^i{}_A{}^{jk}{}_B{}^l = \kappa^3 \bar\psi^i \gamma_A \psi^j \bar\psi^k \gamma_B \psi^l \vert w \vert, 
\ \cdots, 
\label{bosonic0}
\\[1mm]
\A \A 
f^i = \psi^i \vert w \vert, \ f^{ij}{}_A{}^k = \kappa^2 \psi^i \bar\psi^j \gamma_A \psi^k \vert w \vert, 
\ \cdots, 
\label{fermionic0}
\ea
for $n = 1, 2, \cdots$, respectively. 
In the functionals (\ref{bosonic}) and (\ref{fermionic}), 
(Lorentz) indices $A, B, \cdots$ are used as ones for a basis of $\gamma$ matrices in $d = 2$, 
i.e., $\gamma_A = {\bf 1}, \gamma_5 \ {\rm or}\ i \gamma_a$ ($\gamma^A = {\bf 1}, \gamma_5 \ {\rm or}\ -i \gamma^a$). 
Note that vector components appear in the functionals $b^i{}_A{}^j$ of Eq.(\ref{bosonic0}) for $N \ge 2$ SUSY 
and the definitions of the functionals (\ref{bosonic}) and (\ref{fermionic}) (or (\ref{bosonic0}) and (\ref{fermionic0})) 
terminate with $n = N + 1$ and $n = N$, because $(\psi^i)^n = 0$ for $n \ge 2N+1$. 
In the case for $N = 2$ SUSY in $d = 2$, we have already shown that the functionals (\ref{bosonic}) and (\ref{fermionic}) 
lead to the NL/LSUSY relations for a (massless) vector linear supermultiplet with general auxiliary fields 
prior to transforming to $U(1)$ gauge supermultiplets by using the superspace formalism \cite{ST2}. 

Then, the variations of the functionals (\ref{bosonic}) and (\ref{fermionic}) 
under the NLSUSY transformations (\ref{NLSUSY}) and (\ref{NLSUSYdetw}) become 
\ba
\delta_\zeta b^i{}_A{}^{jk}{}_B{}^{l \cdots m}{}_C{}^n 
\A = \A 
\kappa^{2(n-1)} \left[ \left\{ \left( \bar\zeta^i \gamma_A \psi^j + \bar\psi^i \gamma_A \zeta^j \right) 
\bar\psi^k \gamma_B \psi^l \cdots \bar\psi^m \gamma_C \psi^n + \cdots \right\} \vert w \vert \right. 
\nonu
\A \A 
\left. + \kappa \partial_a \left( \xi^a \bar\psi^i \gamma_A \psi^j \bar\psi^k \gamma_B \psi^l 
\cdots \bar\psi^m \gamma_C \psi^n \vert w \vert \right) \right], 
\label{variation1}
\\[1mm]
\delta_\zeta f^{ij}{}_A{}^{kl}{}_B{}^{ml \cdots n}{}_C{}^p
\A = \A 
\kappa^{2n-1} \left[ \left\{ \zeta^i \bar\psi^j \gamma_A \psi^k \bar\psi^l \gamma_B \psi^m \cdots \bar\psi^n \gamma_C \psi^p 
\right. \right. 
\nonu
\A \A 
\left. + \psi^i \left( \bar\zeta^j \gamma_A \psi^k + \bar\psi^j \gamma_A \zeta^k \right) 
\bar\psi^l \gamma_B \psi^m \cdots \bar\psi^n \gamma_C \psi^p + \cdots \right\} \vert w \vert 
\nonu
\A \A 
\left. + \kappa \partial_a \left( \xi^a \psi^i \bar\psi^j \gamma_A \psi^k \bar\psi^l \gamma_B \psi^m 
\cdots \bar\psi^n \gamma_C \psi^p \vert w \vert \right) \right], 
\label{variation2}
\ea
which indicate that the bosonic and fermionic functionals in Eqs.(\ref{bosonic}) and (\ref{fermionic}) 
are linearly transformed to each other.

\section{Commutator-based linearization of NLSUSY in $d = 2$}

In this section, we show general forms of LSUSY transformations of basic components for vector supermultiplets in the $d = 2$ case, 
which are obtained from the variations (\ref{variation1}) and (\ref{variation2}) 
and the commutation relation (\ref{NLSUSYcomm2}). 
We also explain their derivations by reviewing the procedures of the commutator-based linearization \cite{MT1}. 

By using the set of the functionals (\ref{bosonic}) and (\ref{fermionic}), let us define basic bosonic components as 
\ba
\A \A 
D = b(\psi), 
\ \ M^i{}_A{}^j = \alpha_{1A} b^i{}_A{}^j(\psi), 
\ \ C^i{}_A{}^{jk}{}_B{}^l = \alpha_{3AB} b^i{}_A{}^{jk}{}_B{}^l(\psi), 
\nonu
\A \A 
E^i{}_A{}^{jk}{}_B{}^{lm}{}_C{}^n = \alpha_{5ABC} b^i{}_A{}^{jk}{}_B{}^{lm}{}_C{}^n(\psi), \ \ \cdots 
\label{b-comp}
\ea
and fermionic ones as 
\be
\lambda^i = f^i(\psi), 
\ \ \Lambda^{ij}{}_A{}^k = \alpha_{2A} f^{ij}{}_A{}^k(\psi), 
\ \ \Psi^{ij}{}_A{}^{kl}{}_B{}^m = \alpha_{4AB} f^{ij}{}_A{}^{kl}{}_B{}^m(\psi), \ \ \cdots 
\label{f-comp}
\ee
where $\alpha_{mA} \ (m = 1,2,\cdots,5)$ mean constants whose values are determined 
from definitions of fundamental actions in $d = 2$ LSUSY theories and the invariances of the actions 
under LSUSY transformations of the component fields. 

For the components (\ref{b-comp}) and (\ref{f-comp}), 
LSUSY transformations which satisfy the commutator algebra (\ref{NLSUSYcomm}) can be uniquely expressed as follows; 
\ba
\A \A 
\delta_\zeta D = -i \bar\zeta^i \!\!\not\!\partial \lambda^i, 
\label{v-D} \\[1mm]
\A \A 
\delta_\zeta \lambda^i 
= D \zeta^i - {i \over {2 \alpha_{1A}}} \varepsilon_{M1} \gamma^A \!\!\not\!\partial M^i{}_A{}^j \zeta^j, 
\label{v-lambda} \\[1mm]
\A \A 
\delta_\zeta M^i{}_A{}^j 
= \alpha_{1A} \left( \bar\zeta^i \gamma_A \lambda^j + \bar\lambda^i \gamma_A \zeta^j 
+ {i \over {2 \alpha_{2B}}} \varepsilon_{\Lambda1} 
\bar\zeta^k \!\!\not\!\partial \gamma^B \gamma_A \Lambda^{ij}{}_B{}^k \right), 
\label{v-M} \\[1mm]
\A \A 
\delta_\zeta \Lambda^{ij}{}_A{}^k 
= \alpha_{2A} \left\{ {1 \over \alpha_{1A}} M^j{}_A{}^k \zeta^i 
- {1 \over {2 \alpha_{1B}}} \gamma^B \left( \varepsilon_{M2} \gamma_A \varepsilon'_{M2} M^i{}_B{}^k \zeta^j 
+ \gamma_A \varepsilon'_{M2} M^i{}_B{}^j \zeta^k \right) \right. 
\nonu
\A \A 
\hspace{1.8cm} 
\left. + {i \over {4 \alpha_{3ACB}}} \varepsilon_{C1} \varepsilon'_{C1} 
\gamma^B \!\!\not\!\partial C^{iCjk}{}_{ACB}{}^l \zeta^l \right\}, 
\label{v-Lambda} \\[1mm]
\A \A 
\delta_\zeta C^i{}_A{}^{jk}{}_B{}^l 
= \alpha_{3AB} \left\{ {1 \over \alpha_{2B}} \left( \bar\zeta^i \gamma_A \Lambda^{jk}{}_B{}^l 
+ \varepsilon_{\Lambda2} \bar\zeta^j \gamma_A \Lambda^{ik}{}_B{}^l \right) \right. 
\nonu
\A \A 
\hspace{2.2cm} 
\left. - {1 \over {2 \alpha_{2C}}} \varepsilon_{\Lambda2} 
\left( \bar\zeta^k \gamma_B \gamma^C \gamma_A \Lambda^{ij}{}_C{}^l 
+ \varepsilon'_{\Lambda2} \bar\zeta^l \gamma_B \gamma^C \gamma_A \Lambda^{ij}{}_C{}^k \right) \right\}, 
\nonu
\A \A 
\hspace{2.2cm} 
- {{i \alpha_{3AB}} \over {4 \alpha_{4AB}}} \varepsilon_{\Psi1} \varepsilon_{\Psi2} 
\bar\zeta^m \!\!\not\!\partial \gamma^C \gamma_B \gamma^D \gamma_A \Psi^{ij}{}_D{}^{kl}{}_C{}^m, 
\label{v-C}
\ea
$\cdots ,$ etc. 
In Eqs. from (\ref{v-D}) to (\ref{v-C}), we use a sign factor $\varepsilon$ (or $\varepsilon'$) 
which appears from the relation $\bar\psi^j \gamma_A \psi^i = \varepsilon \bar\psi^i \gamma_A \psi^j$ 
\footnote
{The sign factor $\varepsilon$ is $\varepsilon = +1$ for $\gamma_A = {\bf 1}$ 
and $\varepsilon = -1$ for $\gamma_A = \gamma_5, i \gamma_a$. 
}
and we define $C^{iCjk}{}_{ACB}{}^l = \alpha_{3ACB} b^{iCjk}{}_{ACB}{}^l$ 
\ ($= \alpha_{3ACB} \kappa^3 \bar\psi^i \gamma^C \psi^j$ $\bar\psi^k \gamma_A \gamma_C \gamma_B \psi^l \vert w \vert$) 
with constants $\alpha_{3ACB}$ in Eq.(\ref{v-Lambda}) for convenience of the expression, 
which can be expanded by means of the components $C^i{}_A{}^{jk}{}_B{}^l$ under the Clifford algebra for $\gamma$ matrices. 
Note that the determination of LSUSY transformations terminates with those of bosonic components 
for the functionals at ${\cal O}\{ (\psi^i)^{2N} \}$ in Eq.(\ref{bosonic}). 

Let us explain below the derivations of the LSUSY transformations from (\ref{v-D}) to (\ref{v-C}): 
First, starting with the variation of $D(\psi)$ in the bosonic components (\ref{b-comp}), 
the LSUSY transformations of $D$ and $\lambda^i$ are unambiguously determined as Eqs.(\ref{v-D}) and (\ref{v-lambda}) 
by using a Fierz transformation. 
Then, the closure of the commutator algebra on $D$ under the LSUSY transformations (\ref{v-D}) and (\ref{v-lambda}) 
is guaranteed by means of the commutation relation (\ref{NLSUSYcomm2}), though its straightforward calculation is easy. 

Next, as for the LSUSY transformations (\ref{v-M}) and (\ref{v-Lambda}), 
variations of the functionals $M^i{}_A{}^j(\psi)$ and $\Lambda^{ij}{}_A{}^k(\psi)$ in Eqs.(\ref{b-comp}) and (\ref{f-comp}) 
under the NLSUSY transformations (\ref{NLSUSY}) 
are calculated by following Eqs.(\ref{variation1}) and (\ref{variation2}) as 
\ba
\delta_\zeta M^i{}_A{}^j \A = \A \alpha_{1A} \left\{ \bar\zeta^i \gamma_A \lambda^j + \bar\lambda^i \gamma_A \zeta^j 
- i \kappa^2 \partial_a \left( \bar\zeta^k \gamma^a \psi^k \bar\psi^i \gamma_A \psi^j \vert w \vert \right) \right\}, 
\label{v-M0} 
\\[1mm]
\delta_\zeta \Lambda^{ij}{}_A{}^k 
\A = \A \alpha_{2A} \left\{ {1 \over \alpha_{1A}} M^j{}_A{}^k \zeta^i 
- {1 \over {2 \alpha_{1B}}} \gamma^B 
\left( \varepsilon_{M2} \gamma_A \varepsilon'_{M2} M^i{}_B{}^k \zeta^j + \gamma_A \varepsilon'_{M2} M^i{}_B{}^j \zeta^k \right) \right. 
\nonu
\A \A 
\left. - {i \over 2} \kappa^3 \partial_a 
\left( \gamma^B \gamma^a \zeta^l \bar\psi^l \gamma_B \psi^i \bar\psi^j \gamma_A \psi^k \vert w \vert \right) \right\}, 
\label{v-Lambda0}
\ea
respectively. In the variations (\ref{v-M0}) and (\ref{v-Lambda0}), 
deformations of the functionals $\psi^k \bar\psi^i \gamma_A \psi^j$ 
and $\bar\psi^l \gamma_B \psi^i \bar\psi^j \gamma_A \psi^k$ in the last terms 
are problems for the definition of the LSUSY transformations. 
These are solved by examining two supertransformations of $\lambda^i(\psi)$ and $M^i{}_A{}^j(\psi)$ 
which satisfy the commutation relation (\ref{NLSUSYcomm2}): 
Indeed, the two supertransformations of $\lambda^i(\psi)$ are obtained from Eqs.(\ref{v-D}), (\ref{v-lambda}) and (\ref{v-M0}) as 
\ba
\delta_{\zeta_1} \delta_{\zeta_2} \lambda^i 
\A = \A \delta_{\zeta_1} D \ \zeta_2^i 
- {i \over {4 \alpha_{1A}}} \varepsilon_{M1} \gamma^A \!\!\not\!\partial \ \delta_{\zeta_1} M^i{}_A{}^j \ \zeta_2^j 
\nonu
\A = \A i \bar\zeta_1^j \gamma^a \zeta_2^j \partial_a \lambda^i 
+ \left[ (1 \leftrightarrow 2)\ {\rm symmetric\ terms\ of}\ \partial_a \lambda \right] 
\nonu
\A \A 
+ {1 \over 4} \varepsilon_{M1} \kappa^2 \gamma^A \!\!\not\!\partial \partial_a 
\left( \bar\zeta_1^k \gamma_B \zeta_2^j \gamma^B \gamma^a \psi^k \bar\psi^i \gamma_A \psi^j \vert w \vert \right). 
\label{twosuper-lambda}
\ea
Since the last terms in Eq.(\ref{twosuper-lambda}) are symmetric under exchanging the indices $1$ and $2$ 
of the spinor transformation parameters ($\zeta_1^k$, $\zeta_2^j$) 
and they vanish in the commutation relation (\ref{NLSUSYcomm2}), 
the $\psi^k$ and $\psi^j$ have to go into bilinear forms $\bar\psi^j \gamma_A \psi^k$ 
in the last terms of Eq.(\ref{v-M0}) (and Eq.(\ref{twosuper-lambda})) 
in order to realize straightforwardly the symmetries of the indices of the spinor transformation parameters 
in the components $\Lambda^{ij}{}_A{}^k$ of LSUSY theories. 

Thus the LSUSY transformations of $M^i{}_A{}^j$ are uniquely determined from Eq.(\ref{v-M0}) as Eq.(\ref{v-M}) 
by using a Fierz transformation. 
The LSUSY transformations (\ref{v-lambda}) of $\lambda^i$ satisfy the commutator algebra (\ref{NLSUSYcomm}) 
under Eqs.(\ref{v-D}) and (\ref{v-M}). 

In the same way, according to the two supertransformations of $M^i{}_A{}^j(\psi)$ which are given 
from Eqs.(\ref{v-lambda}), (\ref{v-M}) and (\ref{v-Lambda0}) as 
\ba
\delta_{\zeta_1} \delta_{\zeta_2} M^i{}_A{}^j 
\A = \A i \bar\zeta_1^k \gamma^a \zeta_2^k \partial_a M^i{}_A{}^j 
+ \left[ (1 \leftrightarrow 2)\ {\rm symmetric\ terms\ of}\ D \ {\rm and}\ \partial_a M^i{}_A{}^j \right] 
\nonu
\A \A 
+ {1 \over 4} \varepsilon_{\Lambda1} \kappa^3 \partial_a \partial_b 
\left( \bar\zeta_2^k \gamma^a \gamma^B \gamma_A 
\gamma^C \gamma^b \zeta_1^l \bar\psi^l \gamma_C \psi^i \bar\psi^j \gamma_B \psi^k \vert w \vert \right), 
\label{twosuper-M}
\ea
the $\psi^l$ and $\psi^k$ have to take bilinear forms $\bar\psi^k \gamma_A \psi^l$ 
in the last terms of Eqs.(\ref{v-Lambda0}) and (\ref{twosuper-M}) 
in order to realize the symmetries of the indices of the spinor transformation parameters ($\zeta_1^l$, $\zeta_2^k$) 
in the components $C^i{}_A{}^{jk}{}_B{}^l$ of LSUSY theories. 

Therefore, the LSUSY transformations of $\Lambda^{ij}{}_A{}^k$ are defined from Eq.(\ref{v-Lambda0}) 
as Eq.(\ref{v-Lambda}) by using a Fierz transformation 
and then the LSUSY transformations (\ref{v-M}) of $M^i{}_A{}^j$ satisfy the commutator algebra (\ref{NLSUSYcomm}) 
under Eqs.(\ref{v-lambda}) and (\ref{v-Lambda}). 

Furthermore, as an example of the LSUSY transformations of the components for the higher-order functionals 
of $\psi^i$ than $\Lambda^{ij}{}_A{}^k(\psi)$, 
the LSUSY transformations (\ref{v-C}) are determined by estimating the variations of $C^i{}_A{}^{jk}{}_B{}^l(\psi)$, 
\ba
\delta_\zeta C^i{}_A{}^{jk}{}_B{}^l 
\A = \A \alpha_{3AB} \kappa^2 \left[ \left\{ 
\left( \bar\zeta^i \gamma_A \psi^j + \bar\psi^i \gamma_A \zeta^j \right) \bar\psi^k \gamma_B \psi^l 
\right. \right. 
\nonu
\A \A 
\hspace{1.5cm}
\left. \left. 
+ \bar\psi^i \gamma_A \psi^j \left( \bar\zeta^k \gamma_B \psi^l + \bar\psi^k \gamma_B \zeta^l \right) 
\right\} \vert w \vert \right. 
\nonu
\A \A 
\left. - i \kappa^2 \partial_a 
\left( \bar\zeta^m \gamma^a \psi^m \bar\psi^i \gamma_A \psi^j \bar\psi^k \gamma_B \psi^l \vert w \vert \right) \right]. 
\label{v-C0}
\ea
In the variations (\ref{v-C0}), we have to consider deformations of the functionals both at ${\cal O}(\psi^3)$ and at ${\cal O}(\psi^5)$. 

The LSUSY transformations of $C^i{}_A{}^{jk}{}_B{}^l$ to the fermionic components $\Lambda^{ij}{}_A{}^k$ in Eq.(\ref{v-C}) 
are determined from the functionals at ${\cal O}(\psi^3)$ in Eq.(\ref{v-C0}) 
by taking into account derivative terms of $\Lambda^{ij}{}_A{}^k$ ($\partial \Lambda$-terms) 
in two supertransformations of $\Lambda^{ij}{}_A{}^k(\psi)$. 
In fact, the $\partial \Lambda$-terms which are given from the LSUSY transformations (\ref{v-M}) and (\ref{v-Lambda}) become 
\ba
\A \A 
\delta_{\zeta_1} \delta_{\zeta_2} \Lambda^{ij}{}_A{}^k 
\left[ \partial \Lambda \ {\rm terms\ obtained\ through}\ \delta_\zeta M \right] 
\nonu
\A \A 
\hspace{7mm} 
= {i \over 2} \alpha_{2A} \left[ {1 \over \alpha_{2B}} \varepsilon_{\Lambda1} 
\partial_a \left( \bar\zeta_1^l \gamma^a \gamma^B \gamma_A \Lambda^{jk}{}_B{}^l \zeta_2^i \right) \right. 
\nonu
\A \A 
\hspace{1cm}
- {1 \over {2 \alpha_{2C}}} \varepsilon'_{M2} \varepsilon_{\Lambda1} \gamma^B 
\left\{ \varepsilon_{M2} \gamma_A \partial_a \left( \bar\zeta_1^l \gamma^a \gamma^C \gamma_B \Lambda^{ik}{}_C{}^l \zeta_2^j \right) 
\right. 
\nonu
\A \A 
\hspace{4.5cm}
\left. + \gamma_A \partial_a \left( \bar\zeta_1^l \gamma^a \gamma^C \gamma_B \Lambda^{ij}{}_C{}^l \zeta_2^k \right) 
\right\} \bigg], 
\label{twosuper-Lambda}
\ea
where the $\partial_a (\Lambda^{jk}{}_A{}^l, \Lambda^{ik}{}_A{}^l, \Lambda^{ij}{}_A{}^l)$-type terms appear. 
Since these terms in Eq.(\ref{twosuper-Lambda}) cancel with ones which are obtained from 
the LSUSY transformations of $C^i{}_A{}^{jk}{}_B{}^l$ in the commutation relation (\ref{NLSUSYcomm2}) on $\Lambda^{ij}{}_A{}^k$, 
the variations (\ref{v-C0}) at ${\cal O}(\psi^3)$ have to give $\Lambda$-terms with the same arrangement 
of the internal indices as in Eq.(\ref{twosuper-Lambda}).  
Thus the LSUSY transformations of $C^i{}_A{}^{jk}{}_B{}^l$ are defined with respect to the fermionic components $\Lambda^{ij}{}_A{}^k$ 
as in Eq.(\ref{v-C}) from the ${\cal O}(\psi^3)$-terms of Eq.(\ref{v-C0}) by using Fierz transformations. 
Note that the terms for $\Lambda^{ij}{}_C{}^k$ in Eq.(\ref{v-C}) give the translations of 
$\Lambda^{ij}{}_A{}^k$ (i.e., $\Xi^a \partial_a \Lambda^{ij}{}_A{}^k$) 
in a commutator algebra for the LSUSY transformations (\ref{v-Lambda}). 

As for the functionals at ${\cal O}(\psi^5)$ in the variations (\ref{v-C0}), 
if we consider the two supertransformations of $\Lambda^{ij}{}_A{}^k(\psi)$, 
then we can straghtforwardly confirm that second-order derivative terms of those functionals vanish 
in the commutation relation (\ref{NLSUSYcomm2}) on $\Lambda^{ij}{}_A{}^k$, 
provided that the last terms at ${\cal O}(\psi^5)$ in Eq.(\ref{v-C0}) are expressed as 
\be
\delta_\zeta C^i{}_A{}^{jk}{}_B{}^l \left[ \partial_a f^{ki}{}_A{}^{jl}{}_B{}^m \ {\rm terms} \right] 
= {i \over 2} \varepsilon_{\Psi1} \alpha_{3AB} \kappa^4 \partial_a 
\left( \bar\zeta^m \gamma^a \gamma_C \gamma_B \psi^k \bar\psi^i \gamma_A \psi^j 
\bar\psi^l \gamma^C \psi^m \vert w \vert \right). 
\label{v-C1}
\ee
by means of a Fierz transformation. 
This is because the form of the variation (\ref{v-C1}) contain the bilinear terms $\bar\psi^l \gamma^C \psi^m$ with symmetries 
which correspond to the indices of spinor transformation parameters $(\zeta_1^m, \zeta_2^l)$ 
in the two supertransformations of $\Lambda^{ij}{}_A{}^k$ obtained from the LSUSY transformations (\ref{v-Lambda}). 

However, in order to determine the LSUSY transformations of $C^i{}_A{}^{jk}{}_B{}^l$ (as Eq.(\ref{v-C})), 
a deformation of $\psi^k \bar\psi^i \gamma_A \psi^j$ in the functionals of Eq.(\ref{v-C1}) 
is a remaining problem. 
Therefore, we further examine two supertransformations of $C^i{}_A{}^{jk}{}_B{}^l(\psi)$ 
together with the definition of LSUSY transformations of the basic components $\Psi^{ij}{}_A{}^{kl}{}_B{}^m$ 
in the fermionic components (\ref{f-comp}): 
Derivative terms of $C^i{}_A{}^{jk}{}_B{}^l$ ($\partial C$-terms) in two supertransformations of $C^i{}_A{}^{jk}{}_B{}^l(\psi)$, 
which are given through the LSUSY transformations (\ref{v-Lambda}) and $\Lambda$-terms in Eq.(\ref{v-C}) are 
\ba
\A \A 
\delta_{\zeta_1} \delta_{\zeta_2} C^i{}_A{}^{jk}{}_B{}^l 
\left[ \partial C \ {\rm terms\ obtained\ through}\ \delta_\zeta \Lambda \right] 
\nonu
\A \A 
\hspace{7mm} 
= {{i \alpha_{3AB}} \over {16 \alpha_{3ABC}}} \varepsilon_{C1} \varepsilon'_{C1} \left\{ 
\left( \bar\zeta_2^i \gamma_A \gamma^C \!\!\not\!\partial C^{jDkl}{}_{BDC}{}^m \zeta_1^m 
+ \varepsilon_{\Lambda2} \bar\zeta_2^j \gamma_A \gamma^C \!\!\not\!\partial C^{iDkl}{}_{BDC}{}^m \zeta_1^m \right) 
\right. 
\nonu
\A \A 
\hspace{3.5cm} 
- {1 \over 2} \varepsilon_{\Lambda2} 
\left( \bar\zeta_2^k \gamma_B \gamma_C \gamma_A \gamma^D \!\!\not\!\partial C^{iEjlC}{}_{ED}{}^m \zeta_1^m \right. 
\nonu
\A \A 
\hspace{5cm}
\left. + \varepsilon'_{\Lambda2} \bar\zeta_2^l \gamma_B \gamma_C \gamma_A \gamma^D 
\!\!\not\!\partial C^{iEjkC}{}_{ED}{}^m \zeta_1^m \right) \bigg\}, 
\label{twosuper-C}
\ea
where the $\partial_a (C^j{}_A{}^{kl}{}_B{}^m, C^i{}_A{}^{kl}{}_B{}^m, 
C^i{}_A{}^{jl}{}_B{}^m, C^i{}_A{}^{jk}{}_B{}^m)$-type terms appear. 

In accordance with the internal indices of $\partial C$-terms in Eq.(\ref{twosuper-C}), 
LSUSY transformations of $\Psi^{ij}{}_A{}^{kl}{}_B{}^m$ are determined with respect to $C^i{}_A{}^{jk}{}_B{}^l$ 
by using Fierz transformations in the variations of $\Psi^{ij}{}_A{}^{kl}{}_B{}^m(\psi)$ as 
\footnote{
Note that $\Lambda$-terms in the commutation relation (\ref{NLSUSYcomm2}) on $\Psi^{ij}{}_A{}^{kl}{}_B{}^m$ vanish 
only if the LSUSY transformations (\ref{v-Psi}) are defined, 
since the LSUSY ones of $C^i{}_A{}^{jk}{}_B{}^l$ to $\Lambda$-terms have already been determined as Eq.(\ref{v-C}). 
}
\ba
\A \A 
\delta_\zeta \Psi^{ij}{}_A{}^{kl}{}_B{}^m \left[ C \ {\rm terms} \right] 
\nonu
\A \A 
= \alpha_{4AB} \left\{ {1 \over \alpha_{3AB}} \zeta^i C^j{}_A{}^{kl}{}_B{}^m \right. 
\nonu
\A \A 
\hspace{3mm} - {1 \over {2 \alpha_{3CB}}} 
\left( \varepsilon_{C2} \gamma_C \gamma_A \zeta^j \varepsilon'_{C2} C^{iCkl}{}_B{}^m 
+ \gamma_C \gamma_A \zeta^k \varepsilon'_{C2} C^{iCjl}{}_B{}^m \right) 
\nonu
\A \A 
\hspace{3mm} + {1 \over {4 \alpha_{3ADC}}} 
\left( \varepsilon_{C3} \gamma_C \gamma_B \zeta^l \varepsilon'_{C3} \varepsilon''_{C3} C^i{}_D{}^{jk}{}_A{}^{DCm} 
\right. 
\nonu
\A \A 
\hspace{2.6cm}
\left. \left. 
+ \gamma_C \gamma_B \zeta^m \varepsilon'_{C3} \varepsilon''_{C3} C^i{}_D{}^{jk}{}_A{}^{DCl} \right) \right\}. 
\label{v-Psi}
\ea
Then, in the commutation relation (\ref{NLSUSYcomm2}) on $C^i{}_A{}^{jk}{}_B{}^l$, 
the $\partial C$-terms of Eq.(\ref{twosuper-C}) cancel with ones obtained through the LSUSY transformations (\ref{v-Psi}), 
provided that the LSUSY transformations of $C^i{}_A{}^{jk}{}_B{}^l$ are defined with respect to $\Psi^{ij}{}_A{}^{kl}{}_B{}^m$ 
as in Eq.(\ref{v-C}). 
Here we also note that the last terms with respect to $C^i{}_D{}^{jk}{}_A{}^{DCl}$ 
in the LSUSY transformations (\ref{v-Psi}) give the translations of $C^i{}_A{}^{jk}{}_B{}^l$ 
(i.e., $\Xi^a \partial_a C^i{}_A{}^{jk}{}_B{}^l$) 
in a commutator algebra for the LSUSY transformations of $C^i{}_A{}^{jk}{}_B{}^l$ through Eq.(\ref{v-C}). 

Thus the LSUSY transformations of $C^i{}_A{}^{jk}{}_B{}^l$ are uniquely determined as Eq.(\ref{v-C}). 
The LSUSY transformations (\ref{v-Lambda}) of $\Lambda^{ij}{}_A{}^k$ 
satisfy the commutator algebra (\ref{NLSUSYcomm}) under Eqs.(\ref{v-M}) and (\ref{v-C}). 
As for LSUSY transformations of the basic components for higher-order functionals than $C^i{}_A{}^{jk}{}_B{}^l(\psi)$, 
they can be determined in accordance with the arguments for the definition of LSUSY transformations in this section. 

\section{Reduction to a $N = 2$ vector supermultiplet}

In this section, we reduce the basic components (\ref{b-comp}) and (\ref{f-comp}) to the ones for $N = 2$ SUSY 
based on the functional structure of the NG fermions and derive LSUSY transformations for a $N = 2$ vector supermultiplets 
from Eqs.(\ref{v-D}) to (\ref{v-C}) as an instructive example in order to discuss the NL/LSUSY relations for $N \ge 3$ SUSY in $d = 2$. 
For $N = 2$ SUSY, the basic components for the vector supermultiplet are constituted from 
($D$, \ $\lambda^i$, \ $M^i{}_A{}^j$, \ $\Lambda^{ij}{}_A{}^k$, \ $C^i{}_A{}^{jk}{}_B{}^l$) in Eqs.(\ref{b-comp}) and (\ref{f-comp}), 
which are defined from the functionals of $\psi^i \ (i = 1,2)$ up to ${\cal O}(\psi^4)$. 
The bosonic and fermionic d.o.f. of the components for the lower-order functionals of $\psi^i$ (except for $\vert w \vert$), 
i.e. the d.o.f of ($D$, \ $\lambda^i$, \ $M^i{}_A{}^j$) are ($1$, $4$, $6$), respectively, 
where the components $M^i{}_A{}^j$ are decomposed as 
\ba
\A \A 
M^{ij} = M^{(ij)} = \alpha_{11} \kappa \bar\psi^i \psi^j \vert w \vert, 
\nonu
\A \A 
\phi^{ij} = \phi^{[ij]} = \alpha_{12} \kappa \bar\psi^i \gamma_5 \psi^j \vert w \vert, 
\nonu
\A \A 
v_a^{ij} = v_a^{[ij]} = i \alpha_{13} \kappa \bar\psi^i \gamma_a \psi^j \vert w \vert. 
\label{decomp-M}
\ea

As for the components ($\Lambda^{ij}{}_A{}^k$, \ $C^i{}_A{}^{jk}{}_B{}^l$), 
their d.o.f. have to be counted based on the NG-fermion functional structure under identities 
which are connected with Fierz transformations: 
First, the components $\Lambda^{ij}{}_A{}^k$ are decomposed as 
\ba
\A \A 
\Lambda^{ijk} = \Lambda^{i(jk)} = \alpha_{21} \kappa^2 \psi^i \bar\psi^j \psi^k \vert w \vert, 
\nonu
\A \A 
\Lambda_5^{ijk} = \Lambda_5^{i[jk]} = \alpha_{22} \kappa^2 \psi^i \bar\psi^j \gamma_5 \psi^k \vert w \vert, 
\nonu
\A \A 
\Lambda_a^{ijk} = \Lambda_a^{i[jk]} = i \alpha_{23} \kappa^2 \psi^i \bar\psi^j \gamma_a \psi^k \vert w \vert. 
\label{decomp-Lambda}
\ea
The d.o.f. of $\Lambda^{ijk} = \Lambda^{i(jk)}$ for $i = 1,2$ are apparently 12. 
However, from the viewpoint of the NG-fermion functional structure, 4 components vanish as 
\be
\Lambda^{111}(\psi) = 0, \ \Lambda^{222}(\psi) = 0, 
\label{sub-Lambda01}
\ee
and 4 components in $\Lambda^{ijk}$ are related to each other by means of Fierz transformations as 
\be
\Lambda^{112}(\psi) = -{1 \over 2} \Lambda^{211}(\psi), 
\ \ \Lambda^{221}(\psi) = -{1 \over 2} \Lambda^{122}(\psi). 
\label{sub-Lambda02}
\ee
Therefore, the d.o.f. of $\Lambda^{ijk}$ for $N = 2$ SUSY are effectively $12-8=4$. 

Furthermore, the components $\Lambda_5^{i[jk]}(\psi)$ and $\Lambda_a^{i[jk]}(\psi)$ 
are expressed by using the components $(\Lambda^{122}, \Lambda^{211})(\psi)$ in Eq.(\ref{sub-Lambda02}), i.e. 
\ba
\A \A 
\Lambda_5^{112}(\psi) = -{\alpha_{22} \over {2 \alpha_{21}}} \gamma_5 \Lambda^{211}(\psi), 
\ \ \Lambda_5^{221}(\psi) = -{\alpha_{22} \over {2 \alpha_{21}}} \gamma_5 \Lambda^{122}(\psi), 
\nonu
\A \A 
\Lambda_a^{112}(\psi) = -{{i \alpha_{23}} \over {2 \alpha_{21}}} \gamma_a \Lambda^{211}(\psi), 
\ \ \Lambda_a^{221}(\psi) = -{{i \alpha_{23}} \over {2 \alpha_{21}}} \gamma_a \Lambda^{122}(\psi), 
\label{sub-Lambda03}
\ea
so that fermionic auxiliary fields in the $N = 2$ vector supermultiplet are defined 
from the remaining components $(\Lambda^{122}, \Lambda^{211})$ in $\Lambda^{ij}{}_A{}^k$ as 
\be
\Lambda^i = \Lambda^{ijj}. 
\label{N2-Lambda}
\ee

Second, we count the effective d.o.f. of the components $C^i{}_A{}^{jk}{}_B{}^l$ by decomposing them as 
\ba
\A \A 
C^{ijkl} = C^{(ij)(kl)} = \alpha_{31} \kappa^3 \bar\psi^i \psi^j \bar\psi^k \psi^l \vert w \vert, 
\nonu
\A \A 
C_5^{ijkl} = C_5^{[ij](kl)} = \alpha_{32} \kappa^3 \bar\psi^i \gamma_5 \psi^j \bar\psi^k \psi^l \vert w \vert, 
\ \ \tilde C_5^{ijkl} = \tilde C_5^{(ij)[kl]} = \alpha'_{32} \kappa^3 \bar\psi^i \psi^j \bar\psi^k \gamma_5 \psi^l \vert w \vert, 
\nonu
\A \A 
C_a^{ijkl} = C_a^{[ij](kl)} = i \alpha_{33} \kappa^3 \bar\psi^i \gamma_a \psi^j \bar\psi^k \psi^l \vert w \vert, 
\ \ \tilde C_a^{ijkl} = \tilde C_a^{(ij)[kl]} = i \alpha'_{33} \kappa^3 \bar\psi^i \psi^j \bar\psi^k \gamma_a \psi^l \vert w \vert, 
\nonu
\A \A 
C_{55}^{ijkl} = C_{55}^{[ij][kl]} = \alpha_{34} \kappa^3 \bar\psi^i \gamma_5 \psi^j \bar\psi^k \gamma_5 \psi^l \vert w \vert, 
\nonu
\A \A 
C_{a5}^{ijkl} = C_{a5}^{[ij][kl]} = i \alpha_{35} \kappa^3 \bar\psi^i \gamma_a \psi^j \bar\psi^k \gamma_5 \psi^l \vert w \vert, 
\ \ \tilde C_{5a}^{ijkl} = \tilde C_{5a}^{[ij][kl]} = i \alpha'_{35} \kappa^3 \bar\psi^i \gamma_5 \psi^j \bar\psi^k \gamma_a \psi^l \vert w \vert, 
\nonu
\A \A 
C_{ab}^{ijkl} = C_{ab}^{[ij][kl]} = \alpha_{36} \kappa^3 \bar\psi^i \gamma_a \psi^j \bar\psi^k \gamma_b \psi^l \vert w \vert, 
\label{decomp-C}
\ea
but $C^{1122}$ is the only remaining component in Eq.(\ref{decomp-C}) for $N = 2$ SUSY 
based on the NG-fermion functional structure since 
\ba
\A \A 
C^{1212}(\psi) = -{1 \over 2} C^{1122}(\psi), 
\nonu
\A \A 
C_{55}^{1212}(\psi) = {\alpha_{33} \over {2 \alpha_{31}}} C^{1122}(\psi), 
\ \ C_{ab}^{1212}(\psi) = {\alpha_{36} \over {2 \alpha_{31}}} \eta_{ab} C^{1122}(\psi), 
\label{sub-C}
\ea
and all other components vanish by means of Fierz transformations. 
Then, a bosonic auxiliary field in the $N = 2$ vector supermultiplet is defined from $C^{1122}$ as 
\be
C = C^{iijj}, 
\label{N2-C}
\ee
where we identify $C^{1122}$ with $C^{2211}$ from the NG-fermion functional structure 
and adopt the  $SO(2)$ invariant definition. 

Thus the d.o.f. of the reduced bosonic and fermionic components for the $d = 2$, $N = 2$ vector supermultiplet 
are balanced as $8 = 8$ in Eqs.(\ref{decomp-M}), (\ref{N2-Lambda}) and (\ref{N2-C}). 
Then, the LSUSY transformations (\ref{v-D}) to (\ref{v-C}) for $N = 2$ SUSY become 
\ba
\A \A 
\delta_\zeta D = -i \bar\zeta^i \!\!\not\!\partial \lambda^i, 
\label{N2v-D} 
\\[1mm]
\A \A 
\delta_\zeta \lambda^i 
= D \zeta^i - {i \over {2 \alpha_{11}}} \!\!\not\!\partial M^{ij} \zeta^j 
+ {i \over {2 \alpha_{12}}} \epsilon^{ij} \gamma_5 \!\!\not\!\partial \phi \zeta^j 
+ {1 \over {2 \alpha_{13}}} \epsilon^{ij} \gamma^a \!\!\not\!\partial v_a \zeta^j, 
\label{N2v-lambda}
\\[1mm]
\A \A 
\delta_\zeta M^{11} 
= \alpha_{11} \left( 2 \bar\zeta^1 \lambda^1 
- {i \over \alpha_{21}} \bar\zeta^2 \!\!\not\!\partial \Lambda^2 \right), 
\label{N2v-M1}
\\[1mm]
\A \A 
\delta_\zeta M^{22} 
= \alpha_{11} \left( 2 \bar\zeta^2 \lambda^2 
- {i \over \alpha_{21}} \bar\zeta^1 \!\!\not\!\partial \Lambda^1 \right), 
\label{N2v-M2}
\\[1mm]
\A \A 
\delta_\zeta M^{ij} 
= \alpha_{11} \left\{ \bar\zeta^i \lambda^j + \bar\zeta^j \lambda^i 
+ {i \over {2 \alpha_{21}}} \left( \bar\zeta^i \!\!\not\!\partial \Lambda^j 
+ \bar\zeta^j \!\!\not\!\partial \Lambda^i \right) \right\} \ \ (i \not= j), 
\label{N2v-M3}
\\[1mm]
\A \A 
\delta_\zeta \phi 
= \alpha_{12} \epsilon^{ij} \left( \bar\zeta^i \gamma_5 \lambda^j 
- {i \over {2 \alpha_{21}}} \bar\zeta^i \gamma_5 \!\!\not\!\partial \Lambda^j \right), 
\label{N2v-phi}
\\[1mm]
%
%
\A \A 
\delta_\zeta v_a 
= \alpha_{13} \epsilon^{ij} \left( i \bar\zeta^i \gamma_a \lambda^j 
- {1 \over {2 \alpha_{21}}} \bar\zeta^i \!\!\not\!\partial \gamma_a \Lambda^j \right), 
\label{N2v-v}
\\[1mm]
%
%
%
\A \A 
\delta_\zeta \Lambda^i 
= \alpha_{21} \left\{ {1 \over \alpha_{11}} \left( M^{jj} \zeta^i - M^{ij} \zeta^j \right) \right. 
\nonu
\A \A 
\hspace{1.2cm} \left. + {1 \over \alpha_{12}} \epsilon^{ij} \phi \gamma_5 \zeta^j 
- {i \over \alpha_{13}} \epsilon^{ij} v_a \gamma^a \zeta^j 
- {i \over {4 \alpha_{31}}} \!\!\not\!\partial C \zeta^i \right\}, 
\label{N2v-Lambda}
\\[1mm]
\A \A 
\delta_\zeta C = {{4 \alpha_{31}} \over \alpha_{21}} \bar\zeta^i \Lambda^i, 
\label{N2v-C}
\ea
where we define $\phi$ and $v^a$ by means of 
\ba
\A \A 
\phi = {1 \over 2} \epsilon^{ij} \phi^{ij} 
= {1 \over 2} \alpha_{12} \kappa \epsilon^{ij} \bar\psi^i \gamma_5 \psi^j \vert w \vert, 
\nonu
\A \A 
v_a = {1 \over 2} \epsilon^{ij} v_a^{ij} 
= {i \over 2} \alpha_{13} \kappa \epsilon^{ij} \bar\psi^i \gamma_a \psi^j \vert w \vert. 
\label{N2-v}
\ea
The LSUSY transformations (\ref{N2v-D}) to (\ref{N2v-C}) just correspond to the ones obtained 
from a general $N = 2$ superfield in $d = 2$ \cite{DVF,ST6}.

\section{Transforming to $U(1)$ gauge and scalar supermultiplets for $N = 2$ SUSY}

The $N = 2$ vector supermultiplet obtained in Section 4 is transformed into minimal $U(1)$ gauge or scalar supermultiplet 
by means of appropriate recombinations of the functionals of the NG fermions, 
which are defined from the basic component fields ($D$, $\lambda^i$, $M^{ij}$, $\phi$, $v_a$, $\Lambda^i$, $C$) 
in Eqs.(\ref{b-comp}), (\ref{f-comp}), (\ref{decomp-M}), (\ref{N2-Lambda}), (\ref{N2-C}) and (\ref{N2-v}): 
\footnote{
We can multiply the functionals of $\psi^i$ for ($D$, \ $\lambda^i$, \ $M^i{}_A{}^j$, \ $\Lambda^{ij}{}_A{}^k$, \ $C^i{}_A{}^{jk}{}_B{}^l$) 
by a overall constant $\xi$ which determine the magnitude of a vacuum expectation value of the $D$-term, 
but we take the value $\xi = 1$ for simplicity of the discussions. 
}
The $U(1)$ gauge supermultiplet in the $N = 2$ LSUSY theory is derived from component fields 
which are defined by using the same set of the functionals of $\psi^i$ as in the $N = 2$ vector supermultiplet as follows; 
\ba
\A \A 
\tilde \lambda^i(\psi) = \left( \lambda^i - {i \over {2 \alpha_{21}}} \!\!\not\!\partial \Lambda^i \right)(\psi), 
\label{recombi-lambda}
\\[1mm]
\A \A 
\tilde D(\psi) = \left( D - {1 \over {8 \alpha_{31}}} \Box C \right)(\psi), 
\label{recombi-D}
\\[1mm]
\A \A 
M(\psi) = M^{ii}(\psi), 
\ea
in addition to $\phi(\psi)$ and $v_a(\psi)$ in Eq.(\ref{N2-v}). 

Indeed, the definition (\ref{recombi-lambda}) of spinor fields $\tilde \lambda^i$ 
induce LSUSY transformations with a $U(1)$ gauge field strength $F_{ab} = \partial_a v_b - \partial_b v_a$ 
from Eqs.(\ref{N2v-lambda}) and (\ref{N2v-Lambda}) (under the NLSUSY transformations (\ref{NLSUSY})) as 
\be
\delta_\zeta \tilde \lambda^i 
= \tilde D \zeta^i - {i \over {2 \alpha_{11}}} \!\!\not\!\partial M \zeta^i 
+ {i \over \alpha_{12}} \epsilon^{ij} \gamma_5 \!\!\not\!\partial \phi \zeta^j 
- {1 \over {2 \alpha_{13}}} \epsilon^{ij} \epsilon^{ab} F_{ab} \zeta^j. 
\label{N2U1v-lambda}
\ee
%
Then, LSUSY transformations of an auxiliary scalar field $\tilde D$ in the definition (\ref{recombi-D}) 
and the other bosonic component fields ($M$, $\phi$, $v_a$) become 
\ba
\A \A 
\delta_\zeta \tilde D = -i \bar\zeta^i \!\!\not\!\partial \tilde \lambda^i, 
\label{N2U1v-D} 
\\[1mm]
\A \A 
\delta_\zeta M = 2 \alpha_{11} \bar\zeta^i \tilde \lambda^i, 
\label{N2U1v-M}
\\[1mm]
\A \A 
\delta_\zeta \phi = \alpha_{12} \epsilon^{ij} \bar\zeta^i \gamma_5 \tilde \lambda^j, 
\label{N2U1v-phi}
\\[1mm]
\A \A 
\delta_\zeta v_a 
= i \alpha_{13} \epsilon^{ij} \bar\zeta^i \gamma_a \tilde \lambda^j + \partial_a W_\zeta, 
\label{N2U1v-v}
\ea
so that the component fields ($\tilde D$, $\tilde \lambda^i$, $M$, $\phi$, $v_a$) 
constitute the $U(1)$ gauge supermultiplet, 
in which the components $(M^{11}-M^{22},M^{12})$ do not appear. 

Note that a $U(1)$ gauge transformation parameter $W_\zeta$ in the LSUSY transformations (\ref{N2U1v-v}) is 
\be
W_\zeta = -{\alpha_{13} \over \alpha_{21}} \epsilon^{ij} \bar\zeta^i \Lambda^j, 
\label{N2-W}
\ee
which leads to a relation in a commutator algebra for $v_a$ as 
\be
\delta_{\zeta_1} W_{\zeta_2} - \delta_{\zeta_2} W_{\zeta_1} 
= \alpha_{13} \left( {1 \over \alpha_{11}} \epsilon^{ij} \bar\zeta_1^i \zeta_2^j M 
- {2 \over \alpha_{12}} \bar\zeta_1^i \gamma_5 \zeta_2^i \phi 
+ {2i \over \alpha_{13}} \bar\zeta_1^i \gamma^a \zeta_2^i v_a \right). 
\label{N2-Wcomm}
\ee
Therefore, the commutator algebra for $v_a$ under Eqs.(\ref{N2U1v-lambda}), (\ref{N2U1v-v}) and (\ref{N2-Wcomm}) 
does not induce a $U(1)$ gauge transformation term in contrast to the Wess-Zumino gauge \cite{WZ,WB} 
as a result of the linearization of NLSUSY based on the commutator algebra (\ref{NLSUSYcomm}) in the VA NLSUSY theory. 

When a $N = 2$ LSUSY (free) action for the $U(1)$ gauge supermultiplet, 
\be
S_{N = 2\ {\rm gauge}} = \int d^2x \left\{ 
-{1 \over 4} (F_{ab})^2 + {i \over 2} \bar{\tilde \lambda}^i \!\!\not\!\partial \tilde \lambda^i 
+ {1 \over 2} (\partial_a M)^2 + {1 \over 2} (\partial_a \phi)^2 
+ {1 \over 2} \tilde D^2 - {1 \over \kappa} \tilde D \right\}, 
\label{N2U1action}
\ee
is defined, the values of the constants ($\alpha_{11}$, $\alpha_{12}$, $\alpha_{13}$) are determined 
as $\displaystyle{\alpha_{11}^2 = {1 \over 4}}$ and $\alpha_{12}^2 = \alpha_{13}^2 = 1$ 
from the invariance of the action (\ref{N2U1action}) under the LSUSY transformations 
from (\ref{N2U1v-lambda}) to (\ref{N2U1v-v}). 
In the action (\ref{N2U1action}), the relative scales of the terms for the auxiliary fields $\Lambda$ and $C$ 
to the kinetic terms of the physical fields are fixed by taking the values of ($\alpha_{21}$, $\alpha_{31}$) 
in the recombinations (\ref{recombi-lambda}) and (\ref{recombi-D}), 
e.g. as $\displaystyle{\alpha_{21} = -{1 \over 2}}$ and $\displaystyle{\alpha_{31} = -{1 \over 8}}$. 

The above results for the $U(1)$ gauge supermultiplet in $d = 2$, $N = 2$ LSUSY theory, 
which are obtained from the commutator-based linearization of NLSUSY, 
coincide with the ones in Ref.\cite{ST2} based on the superspace formalism. 
In addition, the relation between the LSUSY action (\ref{N2U1action}) 
and the VA NLSUSY action (\ref{NLSUSYaction}) for $N = 2$ SUSY, 
\be
S_{N = 2\ {\rm gauge}}(\psi) + [\ {\rm a\ surface\ term}\ ] = S_{N = 2\ {\rm NLSUSY}} 
\label{U(1)-NLrelation}
\ee
can be shown, e.g. by means of the superspace formalism (for example, see \cite{ST2,ST7}). 

On the other hand, the scalar supermutiplet in the $N = 2$ LSUSY theory is also derived 
by defining component fields from the same set of the functionals of $\psi^i$ in the $N = 2$ vector supermultiplet as 
\ba
\A \A 
\chi^i(\psi) = \left( \lambda^i + {i \over {2 \alpha_{21}}} \!\!\not\!\partial \Lambda^i \right)(\psi), 
\label{recombi-chi}
\\[1mm]
\A \A 
F(\psi) = \left( D + {1 \over {8 \alpha_{31}}} \Box C \right)(\psi), 
\ \ G(\psi) = \partial^a v_a, 
\label{recombi-FG}
\\[1mm]
\A \A 
A(\psi) = (M^{11} - M^{22})(\psi), 
\ \ B(\psi) = 2 M^{12}(\psi), 
\label{recombi-AB}
\ea
where the components ($M^{11}-M^{22}$, $M^{12}$), which are the redundant ones 
in the LSUSY transformations from (\ref{N2U1v-lambda}) to (\ref{N2U1v-v})
for the $U(1)$ gauge supermultiplet, appear in the definition (\ref{recombi-AB}) of the components ($A$, $B$). 
In fact, the variations of $\chi^i$ in Eq.(\ref{recombi-chi}) under the NLSUSY transformations (\ref{NLSUSY}) 
induce their LSUSY transformations, 
\ba
\A \A 
\delta_\zeta \chi^1 
= F \zeta^1 + {1 \over \alpha_{13}} G \zeta^2 - {i \over {2 \alpha_{11}}} \!\!\not\!\partial (A \zeta^1 + B \zeta^2), 
\nonu
\A \A 
\delta_\zeta \chi^2 
= F \zeta^2 - {1 \over \alpha_{13}} G \zeta^1 + {i \over {2 \alpha_{11}}} \!\!\not\!\partial (A \zeta^2 - B \zeta^1), 
\label{N2Sv-lambda}
\ea
which are also obtained from Eqs.(\ref{N2v-lambda}) and (\ref{N2v-Lambda}). 
Then, LSUSY transformations of the bosonic fields ($F$,$G$,$A$,$B$) in Eqs.(\ref{recombi-FG}) and (\ref{recombi-AB}) become 
\ba
\A \A 
\delta_\zeta F = -i \bar\zeta^i \!\!\not\!\partial \chi^i, 
\ \ \delta_\zeta G = -i \alpha_{13} \epsilon^{ij} \bar\zeta^i \!\!\not\!\partial \chi^j, 
\label{N2Sv-FG} 
\\[1mm]
\A \A 
\delta_\zeta A = 2 \alpha_{11} \left( \bar\zeta^1 \chi^1 - \bar\zeta^2 \chi^2 \right), 
\label{N2Sv-A}
\\[1mm]
\A \A 
\delta_\zeta B = 2 \alpha_{11} \left( \bar\zeta^1 \chi^2 + \bar\zeta^2 \chi^1 \right). 
\label{N2Sv-B}
\ea
Thus the component fields ($F$, $G$, $\chi^i$, $A$, $B$) constitute the scalar supermultiplet. 

When we define a $N = 2$ LSUSY (free) action for the scalar supermultiplet as 
\be
S_{N = 2\ {\rm scalar}} = \int d^2x \left\{ 
{1 \over 2} (\partial_a A)^2 + {1 \over 2} (\partial_a B)^2 
+ {i \over 2} \bar\chi^i \!\!\not\!\partial \chi^i 
+ {1 \over 2} (F^2 + G^2) - {1 \over \kappa} F \right\}, 
\label{N2Saction}
\ee
the values of the constants ($\alpha_{11}$, $\alpha_{13}$) in the LSUSY transformations (\ref{N2Sv-lambda}) to (\ref{N2Sv-B}) 
are determined as $\displaystyle{\alpha_{11}^2 = {1 \over 4}}$ and $\alpha_{13}^2 = 1$ 
from the LSUSY invariance of the action (\ref{N2U1action}). 
In the action (\ref{N2Saction}), the relative scales of the terms for the auxiliary fields $\Lambda$ and $C$ 
to the kinetic terms of the physical fields are fixed by taking the values of ($\alpha_{21}$, $\alpha_{31}$) 
in the recombinations (\ref{recombi-chi}) and (\ref{recombi-FG}), 
e.g. as $\displaystyle{\alpha_{21} = {1 \over 2}}$ and $\displaystyle{\alpha_{31} = {1 \over 8}}$. 
It is expected that the LSUSY action (\ref{N2Saction}) and the VA NLSUSY action (\ref{NLSUSYaction}) for $N = 2$ SUSY 
are related to each other from the NL/LSUSY relation in the $d = 4$, $N = 1$ SUSY theory \cite{IK,Ro}. 

We note that the recombinations of the functionals of the NG fermions in Eqs.(\ref{recombi-lambda}) and (\ref{recombi-D}) 
for the $U(1)$ gauge supermultiplet 
and Eqs.(\ref{recombi-chi}) and (\ref{recombi-FG}) for the scalar supermultiplet 
by using the auxiliary component fields $\Lambda^i(\psi)$ and $C(\psi)$ (and $v_a(\psi)$) 
have the same form as the ones for $N = 1$ NL/LSUSY relations in $d = 4$ \cite{MT2}.

\section{Structure of basic components for a $N = 3$ vector supermultiplet}

In this section, we focus on a $N = 3$ vector supermultiplet derived from the VA NLSUSY theory 
as the simplest but a {\it general} example of the NL/LSUSY relations in $d = 2$ 
and explain structures of the basic components (\ref{b-comp}) and (\ref{f-comp}) for $N = 3$ SUSY 
by counting their bosonic and fermionic d.o.f. based on the NG-fermion functionals in Eqs.(\ref{bosonic}) and (\ref{fermionic}). 
From the consideration of the reduction of the basic components (\ref{b-comp}) and (\ref{f-comp}) 
to the ones for the $N = 2$ vector supermultiplet in Section 4, 
let us give two procedures for counting the d.o.f. of the components for vector supermultiplets 
in $N$-extended SUSY based on the NLSUSY phase as follows: \\[1mm]
(a) \ The d.o.f. of the basic components are counted based on identities for the NG fermions $\psi^i$, 
which are obtained from Fierz transformations, 
e.g. as in the relations from (\ref{sub-Lambda01}) to (\ref{sub-Lambda03}) and (\ref{sub-C}) for $N = 2$ SUSY. \\[1mm]
(b) \ In order to give the second procedure, 
we define the basic components expressed as the following bosonic or fermionic functionals of $\psi^i$, 
\ba
b^{iijj \cdots kk} \left( (\psi^i)^{2(n-1)} \vert w \vert \right) 
\A = \A \kappa^{2n-3} \bar\psi^i \psi^i \bar\psi^j \psi^j \cdots \bar\psi^k \psi^k \vert w \vert, 
\label{bosonic1} 
\\[1mm]
f^{liijjkk} \left( (\psi^i)^{2n-1} \vert w \vert \right) 
\A = \A \kappa^{2(n-1)} \psi^l \bar\psi^i \psi^i \bar\psi^j \psi^j \cdots \bar\psi^k \psi^k \vert w \vert 
\label{fermionic1}
\ea
with $i \not= j$, $j \not= k$, $k \not= i$. In  Eqs.(\ref{bosonic1}) and (\ref{fermionic1}), 
we consider $n \ge 3$ and use a notation without tensor-contraction rule for simplicity of the equations, 
i.e. they mean 
\ba
\A \A 
b^{1122} = \kappa^3 \bar\psi^1 \psi^1 \bar\psi^2 \psi^2 \vert w \vert, 
\ b^{2211} = \kappa^3 \bar\psi^1 \psi^1 \bar\psi^2 \psi^2 \vert w \vert, 
\nonu
\A \A 
b^{112233} = \kappa^3 \bar\psi^1 \psi^1 \bar\psi^2 \psi^2 \bar\psi^3 \psi^3 \vert w \vert, 
\ b^{113322} = \kappa^3 \bar\psi^1 \psi^1 \bar\psi^3 \psi^3 \bar\psi^2 \psi^2 \vert w \vert, 
\ \cdots, 
\\[1mm]
\A \A 
f^{12233} = \kappa^4 \psi^1 \bar\psi^2 \psi^2 \bar\psi^3 \psi^3 \vert w \vert, 
\ f^{13322} = \kappa^4 \psi^1 \bar\psi^3 \psi^3 \bar\psi^2 \psi^2 \vert w \vert, 
\ \cdots, 
\ea
etc. Then, from the NG-fermion functional structure, 
we identify the each component in the functionals (\ref{bosonic1}) or (\ref{fermionic1}) with the other ones, 
i.e. $b^{1122} = b^{2211}$, $b^{112233} = b^{113322} = \cdots$ and $f^{12233} = f^{13322}$, $\cdots$ etc. as the second procedure. 
This is adopted in the manifest $SO(N)$ covariant (invariant) definition of basic component fields, 
which is same as that of the component $C$ for $N = 2$ SUSY in Eq.(\ref{N2-C}). 

By using the above procedures (a) and (b), let us count the d.o.f. of the basic components for the $N = 3$ vector supermultiplet, 
which are constituted from ($D$, \ $\lambda^i$, \ $M^i{}_A{}^j$, \ $\Lambda^{ij}{}_A{}^k$, \ $C^i{}_A{}^{jk}{}_B{}^l$, 
\ $\Psi^{ij}{}_A{}^{kl}{}_B{}^m$, \ $E^i{}_A{}^{jk}{}_B{}^{lm}{}_C{}^n$)$(\psi)$ 
\ $(i,j,\cdots = 1,2,3)$ in Eqs.(\ref{b-comp}) and (\ref{f-comp}): 
First, the d.o.f. of the components ($D$, \ $\lambda^i$, \ $M^i{}_A{}^j$) are counted (1, 6, 15), straightforwardly, 
where $M^i{}_A{}^j$ are decomposed in Eq.(\ref{decomp-M}). 

Next, as for the fermionic components $\Lambda^{ij}{}_A{}^k$ which are decomposed as Eq.(\ref{decomp-Lambda}), 
for example, the d.o.f. of $\Lambda^{ijk} = \Lambda^{i(jk)}$ are apparently 36. 
However, according to the procedure (a), 6 components in $\Lambda^{ijk}(\psi)$ vanish as 
\be
\Lambda^{iii}(\psi) = (\Lambda^{111}, \Lambda^{222}, \Lambda^{333})(\psi) = 0, 
\label{sub-Lambda1}
\ee
where we use the same component notation as in Eqs.(\ref{bosonic1}) and (\ref{fermionic1}) 
and this notation's rule is also used below in this section. 
In addition, 12 components in $\Lambda^{ijk}(\psi)$ are related to each other by means of Fierz transformations as 
\be
\Lambda^{iij}(\psi) = -{1 \over 2} \Lambda^{jii}(\psi) \ \ (i \not= j). 
\label{sub-Lambda2}
\ee
When we subtract the d.o.f. in Eqs.(\ref{sub-Lambda1}) and (\ref{sub-Lambda2}), the d.o.f. of $\Lambda^{ijk}$ are $36-18=18$. 

In the same way, the d.o.f. of $\Lambda_5^{ijk} = \Lambda_5^{i[jk]}(\psi)$ and $\Lambda_a^{ijk} = \Lambda_a^{i[jk]}(\psi)$ are 
counted as $18-12=6$ and $36-24=12$, respectively, where the following relations, 
\be
\Lambda_5^{iij}(\psi) = -{\alpha_{22} \over {2 \alpha_{21}}} \gamma_5 \Lambda^{jii}(\psi), 
\ \ \Lambda_a^{iij}(\psi) = -{{i \alpha_{23}} \over {2 \alpha_{21}}} \gamma_a \Lambda^{jii}(\psi) \ \ (i \not= j), 
\label{sub-Lambda3}
\ee
are used for the subtraction of the d.o.f. 

Furthermore, we notice that there are relations of linear combinations for the remaining components 
$\Lambda^{ij}{}_A{}^k(\psi)$ ($i \not= j$, $j \not= k$, $k \not= i$) from Fierz transformations as follows; 
\ba
\A \A 
\left( {1 \over \alpha_{21}} a \Lambda^{ijk} + {1 \over \alpha_{22}} b \gamma_5 \Lambda_5^{ijk} 
+ {i \over \alpha_{23}} c \gamma^a \Lambda_a^{ijk} \right)(\psi) 
\nonu
\A \A 
\hspace{1.5cm} = \left( {1 \over \alpha_{21}} a' \Lambda^{kij} + {1 \over \alpha_{22}} b' \gamma_5 \Lambda_5^{kij} 
+ {i \over \alpha_{23}} c' \gamma^a \Lambda_a^{kij} \right)(\psi), 
\label{linear-combi}
\ea
where coefficients $a'$, $b'$, $c'$ are written in terms of $a$, $b$, $c$ as 
\be
a' = -{1 \over 2}(a+b-2c), \ \ b' = {1 \over 2}(a+b+2c), \ \ c' = -{1 \over 2}(a-b). 
\ee
The relations (\ref{linear-combi}) connect the components $\Lambda^{12}{}_A{}^3$ 
with $\Lambda^{31}{}_A{}^2$ and $\Lambda^{23}{}_A{}^1$ in the NLSUSY phase, 
so that the 4 d.o.f. for the two independent relations in Eq.(\ref{linear-combi}) are subtracted from those of $\Lambda^{ij}{}_A{}^k$. 
Therefore, the total d.o.f. of $\Lambda^{ij}{}_A{}^k$ become $(18+6+12)-4=32$ based on the NG-fermion functional structure 
of the components. 

As for the bosonic components $C^i{}_A{}^{jk}{}_B{}^l$ which are decomposed as Eq.(\ref{decomp-C}), 
for example, the d.o.f. of $C^{ijkl} = C^{(ij)(kl)}$ are apparently 36. 
However, according to the procedure (a) based on the NLSUSY phase, 
we have to subtract appropriate d.o.f. from the apparent ones as in the case of the components $\Lambda^{ij}{}_A{}^k$. 
Indeed, 15 components in $C^{ijkl}(\psi)$ are vanish as 
\be
C^{iiii}(\psi) = 0, \ \ C^{iiij}(\psi) = C^{ijii}(\psi) = 0 \ \ (i \not= j). 
\label{sub-C1}
\ee
In addition, 12 components in $C^{ijkl}(\psi)$ are related to each other by means of Fierz transformations as 
\ba
\A \A 
C^{ijij}(\psi) = -{1 \over 2} C^{iijj}(\psi), 
\ \ C^{ijik}(\psi) = -{1 \over 2} \left( d C^{iijk} + e C^{jkii} \right) (\psi) 
\nonu
\A \A
\hspace{7.5cm} (i \not= j, j \not= k, k \not= i). 
\label{sub-C2}
\ea
with coefficients $d$ and $e$. 
Note that we regard $C^{jjii}$ as $C^{iijj}$ in the first relation of Eq.(\ref{sub-C2}) 
because of the procedure (b) for counting the d.o.f. of the components. 
Therefore, the effective d.o.f. of $C^{ijkl}$ become $36-(15+12)=9$. 

In the same way, if we use relations 
for the components $C_5^{ijkl} = C_5^{[ij](kl)}$ and $\tilde C_5^{ijkl} = \tilde C_5^{(ij)[kl]}$, 
\ba
\A \A 
C_5^{ijii}(\psi) = \tilde C_5^{iiij}(\psi) = 0, 
\ \ C_5^{ijij}(\psi) = \tilde C_5^{ijij}(\psi) = 0, 
\nonu
\A \A 
C_5^{ijik}(\psi) = -{\alpha_{32} \over {2 \alpha_{31}}} C_5^{jkii}(\psi), 
\ \ \tilde C_5^{ijik}(\psi) = -{\alpha'_{32} \over {2 \alpha_{31}}} C_5^{iijk}(\psi) 
\nonu
\A \A
\hspace{6.5cm} (i \not= j, j \not= k, k \not= i), 
\label{sub-C4}
\ea
and ones for the components $C_a^{ijkl} = C_a^{[ij](kl)}$ and $\tilde C_a^{ijkl} = \tilde C_a^{(ij)[kl]}$, 
\ba
\A \A 
C_a^{ijii}(\psi) = \tilde C_a^{iiij}(\psi) = 0, 
\ \ C_a^{ijij}(\psi) = \tilde C_a^{ijij}(\psi) = 0, 
\nonu
\A \A 
C_a^{ijik}(\psi) = -{\alpha_{32} \over {2 \alpha_{31}}} C_a^{jkii}(\psi), 
\ \ \tilde C_a^{ijik}(\psi) = -{\alpha'_{32} \over {2 \alpha_{31}}} C_a^{iijk}(\psi) 
\nonu
\A \A
\hspace{6.5cm} (i \not= j, j \not= k, k \not= i), 
\label{sub-C6}
\ea
then the remaining components in $C_5^{ijkl}$, $\tilde C_5^{ijkl}$, $C_a^{ijkl}$ and $\tilde C_a^{ijkl}$ are only 
\ba
\A \A 
\left( C_5^{1233}, \ C_5^{2311}, \ C_5^{3122} \right), 
\ \ \left( \tilde C_5^{3312}, \ \tilde C_5^{1123}, \ \tilde C_5^{2231} \right), 
\nonu
\A \A 
\left( C_a^{1233}, \ C_a^{2311}, \ C_a^{3122} \right), 
\ \ \left( \tilde C_a^{3312}, \ \tilde C_a^{1123}, \ \tilde C_a^{2231} \right) 
\ea
whose d.o.f. are totally 18. 

The d.o.f. of the other components in Eq.(\ref{decomp-C}), 
i.e. those of the latter ones ($C_{55}^{ijkl}$, $C_{a5}^{ijkl}$, $\tilde C_{5a}^{ijkl}$, $C_{ab}^{ijkl}$)$(\psi)$ 
are effectively 0 since all nonvanishing components in them are expressed 
in terms of the former ones ($C^{ijkl}$, $C_5^{ijkl}$, $\tilde C_5^{ijkl}$, $C_a^{ijkl}$, $\tilde C_a^{ijkl}$)$(\psi)$ 
by meas of Fierz transformations in the case of the $N = 3$ SUSY theory. 
Thus, the total d.o.f. of $C^i{}_A{}^{jk}{}_B{}^l$ become $9+18=27$. 

As for the components $\Psi^{ij}{}_A{}^{kl}{}_B{}^m$ and $E^i{}_A{}^{jk}{}_B{}^{lm}{}_C{}^n$, 
in accordance with the procedure (a) and (b) for counting the d.o.f., they are expressed by means of the following components, 
\be
\Psi^i(\psi) = \Psi^{ijjkk}(\psi), \ E(\psi) = E^{iijjkk}(\psi), 
\ee
respectively, whose d.o.f. are 6 and 1. 

Therefore, the d.o.f. of the basic components 
($D$, \ $\lambda^i$, \ $M^i{}_A{}^j$, \ $\Lambda^{ij}{}_A{}^k$, \ $C^i{}_A{}^{jk}{}_B{}^l$, 
\ $\Psi^{ij}{}_A{}^{kl}{}_B{}^m$, \ $E^i{}_A{}^{jk}{}_B{}^{lm}{}_C{}^n$) for $N = 3$ SUSY 
are summarized as (1, 6, 15, 32, 27, 6, 1), respectively, 
in which the d.o.f. of the general bosonic and fermionic components are balanced as $44 = 44$. 
Namely, their functional structure of the NG fermions $\psi^i$ also determines the effective and general d.o.f. 
of the component fields in the $N = 3$ vector supermultiplet.

\section{Transforming to a $U(1)$ gauge supermultiplet for $N = 3$ SUSY}

Helicity states for physical fields in the $N = 3$ LSUSY theory are 
\be
\left[ \ {\bf \underline{1}} (1), 
{\bf \underline{3}} \left( {1 \over 2} \right), 
{\bf \underline{3}} (0), {\bf \underline{1}} \left( -{1 \over 2} \right) \ \right] 
+ \left[ \ {\rm CPT\ conjugate} \ \right], 
\label{N3helicity}
\ee
where ${\bf \underline{n}}(\lambda)$ means the dimension and the helicity 
in the irreducible representation of $SO(3)$ super-Poincar\'e algebra. 
Basic component fields for the helicity states (\ref{N3helicity}) are obtained 
by multiplying the components (\ref{b-comp}) and (\ref{f-comp}) for $N = 3$ SUSY by overall constants $\xi^i$ 
which correspond to vacuum expectation values of $D$-terms in the minimal $U(1)$ gauge supermultiplet \cite{ST0} 
as follows; 
\footnote{
The d.o.f. of the bosonic and fermionic components in Eq.(\ref{N3comp}) are balanced as $132 = 132$ 
from the arguments in Section 6. 
}
\ba
\A \A 
D^i = \xi^i D(\psi), 
\ \ \lambda^{ij} = \xi^i \lambda^j(\psi), 
\nonu
\A \A 
M^{ijk} = M^{i(jk)} = \xi^i M^{jk}(\psi), 
\ \ \phi^{ijk} = \phi^{i[jk]} = \xi^i \phi^{jk}(\psi), 
\ \ v_a^{ijk} = v_a^{i[jk]} = \xi^i v_a^{jk}(\psi) 
\nonu
\A \A 
\Lambda^{ijk}{}_A{}^l = \xi^i \Lambda^{jk}{}_A{}^l(\psi), 
\ \ C^{ij}{}_A{}^{kl}{}_B{}^m = \xi^i C^j{}_A{}^{kl}{}_B{}^m(\psi), 
\nonu
\A \A 
\Psi^{ijk}{}_A{}^{lm}{}_B{}^n = \xi^i \Psi^{jk}{}_A{}^{lm}{}_B{}^n(\psi), 
\ \ E^{ij}{}_A{}^{kl}{}_B{}^{mn}{}_C{}^p = \xi^i E^j{}_A{}^{kl}{}_B{}^{mn}{}_C{}^p(\psi), 
\label{N3comp}
\ea
where we use the decomposition of the components $M^i{}_A{}^j$ in Eq.(\ref{decomp-M}) 
and ($\Lambda^{ij}{}_A{}^k$, $C^i{}_A{}^{jk}{}_B{}^l$) are also decomposed as Eqs.(\ref{decomp-Lambda}) and (\ref{decomp-C}). 
In this section, we argue a transition from the general component fields (\ref{N3comp}) 
to minimal ones for the $U(1)$ gauge supermultiplet. 

Let us start with a consideration of the $U(1)$ gauge invariance in LSUSY transformations 
of the triplet spinor component fields in Eq.(\ref{N3helicity}) by defining them as 
\be
\lambda^i = \epsilon^{ijk} \lambda^{jk}(\psi) 
\label{lambda-tri0}
\ee
and their appropriate functional-recombinations 
in terms of the auxiliary spinor components $\Lambda^{ijk}{}_A{}^l(\psi)$ in Eq.(\ref{N3comp}): 
Namely, we consider the most general form for the recombinations of $\lambda^i(\psi)$ and $\Lambda^{ijk}{}_A{}^l(\psi)$ as 
\ba
\tilde \lambda^i(\psi) \A = \A \left[ \lambda^i + \epsilon^{ijk} 
\left\{ {i \over \alpha_{21}} \!\!\not\!\partial \left( a \Lambda^{jkll} + b \Lambda^{ljkl} \right) 
+ {{i c} \over \alpha_{22}} \gamma_5 \!\!\not\!\partial \Lambda_5^{ljkl} \right. \right. 
\nonu
\A \A 
\left. \left. + {1 \over \alpha_{23}} \left( d \ \partial^a \Lambda_a^{ljkl} 
+ e \epsilon^{ab} \gamma_5 \partial_a \Lambda_b^{ljkl} \right) \right\} \right] (\psi) 
+ {\cal O}(\psi^5), 
\label{tilde-lambda0}
\ea
with constants $(a,b,c,d,e)$, and we determine relations among the constants from the $U(1)$ gauge invariance 
in LSUSY transformations of the spinor fields (\ref{tilde-lambda0}). 

By using the LSUSY transformations of $\lambda^{ij}$ and $\Lambda^{ijk}{}_A{}^l$, 
\ba
\delta_\zeta \lambda^{ij} 
\A = \A D^i \zeta^j - {i \over {2 \alpha_{11}}} \!\!\not\!\partial M^{ijk} \zeta^k 
+ {i \over {2 \alpha_{12}}} \gamma_5 \!\!\not\!\partial \phi^{ijk} \zeta^k 
+ {1 \over {2 \alpha_{13}}} \gamma^a \!\!\not\!\partial v_a^{ijk} \zeta^k, 
\label{N3v-lambda}
\\[1mm]
\delta_\zeta \Lambda^{ijkl} 
\A = \A \alpha_{21} \left\{ {1 \over {2 \alpha_{11}}} (2M^{ikl} \zeta^j - M^{ijk} \zeta^l - M^{ijl} \zeta^k) \right. 
+ {1 \over {2 \alpha_{12}}} (\phi^{ijk} \gamma_5 \zeta^l + \phi^{ijl} \gamma_5 \zeta^k) 
\nonu
\A \A 
\left. - {i \over {2 \alpha_{13}}} (v_a^{ijk} \gamma^a \zeta^l + v_a^{ijl} \gamma^a \zeta^k) \right\} 
+ {\cal O}(C(\psi)), 
\label{N3v-Lambda1}
\\[1mm]
\delta_\zeta \Lambda_5^{ijkl} 
\A = \A \alpha_{22} \left\{ -{1 \over {2 \alpha_{11}}} (M^{ijk} \gamma_5 \zeta^l - M^{ijl} \gamma_5 \zeta^k) 
+ {1 \over {2 \alpha_{12}}} (2 \phi^{ikl} \zeta^j + \phi^{ijk} \zeta^l - \phi^{ijl} \zeta^k) \right. 
\nonu
\A \A 
\left. + {i \over {2 \alpha_{13}}} (v_a^{ijk} \gamma_5 \gamma^a \zeta^l - v_a^{ijl} \gamma_5 \gamma^a \zeta^k) \right\} 
+ {\cal O}(C(\psi)), 
\label{N3v-Lambda2}
\\[1mm]
\delta_\zeta \Lambda_a^{ijkl} 
\A = \A \alpha_{23} \left\{ -{i \over {2 \alpha_{11}}} (M^{ijk} \gamma_a \zeta^l - M^{ijl} \gamma_a \zeta^k) 
+ {i \over {2 \alpha_{12}}} (\phi^{ijk} \gamma_5 \gamma_a \zeta^l - \phi^{ijl} \gamma_5 \gamma_a \zeta^k) \right. 
\nonu
\A \A 
\left. + {1 \over {2 \alpha_{13}}} (2 v_a^{ikl} \zeta^j + v_b^{ijk} \gamma^b \gamma_a \zeta^l - v_b^{ijl} \gamma^b \gamma_a \zeta^k) \right\} 
+ {\cal O}(C(\psi)), 
\label{N3v-Lambda3}
\ea
we can estimate $v_a$-terms in the variations of $\tilde \lambda^i$ in Eq.(\ref{tilde-lambda0}), which are written as 
\ba
\delta \tilde \lambda^i[v_a \ {\rm terms}] 
\A = \A {1 \over {2 \alpha_{13}}} \left[ \epsilon^{jkl} \left\{ \left( - {1 \over 2} + a \right) 
\epsilon^{ab} \partial_a v_b^{jkl} \gamma_5 \zeta^i 
+ \left( {1 \over 2} + a \right) \partial^a v_a^{jkl} \zeta^i \right\} \right. 
\nonu
\A \A 
+ \epsilon^{ijk} \left\{ \left( {1 \over 2} - a + b + c - d + e \right) 
\epsilon^{ab} \partial_a v_b^{ljk} \gamma_5 \zeta^l \right. 
\nonu
\A \A 
+ \left( -{1 \over 2} - a + b + c + d - e \right) \partial^a v_a^{ljk} \zeta^l 
\nonu
\A \A 
+ (b - c + d - 3e) \epsilon^{ab} \partial_a v_b^{ljl} \gamma_5 \zeta^k 
\nonu
\A \A 
+ (b - c - 3d + e) \partial^a v_a^{ljl} \gamma_5 \zeta^k \Big\} \Big]. 
\label{v-lambda(v)}
\ea
In order to make this variations $U(1)$ gauge invariant, 
the vanishments of the terms for $\partial^a v_a^{jkl}$, $\partial^a v_a^{ljk}$, 
$\epsilon^{ijk} \epsilon^{ab} \partial_a v_b^{ljk}$ and $\epsilon^{ijk} \epsilon^{ab} \partial_a v_b^{ljl}$ 
in Eq.(\ref{v-lambda(v)}) are required, i.e. 
\ba
\A \A 
{1 \over 2} + a = 0, 
\nonu
\A \A 
-{1 \over 2} - a + b + c + d - e = 0, 
\nonu
\A \A 
{1 \over 2} - a + b + c - d + e = 0, 
\nonu
\A \A 
b - c + d - 3e = 0, 
\label{U1condition}
\ea
which mean $\displaystyle{a = {-{1 \over 2}}}$, $\displaystyle{c=-b-{1 \over 2}}$, $d=b+1$ and $\displaystyle{e=b+{1 \over 2}}$. 
Then, the variations (\ref{v-lambda(v)}) become 
\be
\delta \tilde \lambda^i[v_a \ {\rm terms}] 
= -{1 \over \alpha_{13}} \left( {1 \over 4} \epsilon^{ab} F_{ab} \gamma_5 \zeta^i 
+ \epsilon^{ijk} \partial^a v_a^{ljl} \gamma_5 \zeta^k \right), 
\label{U1v-lambda(v)}
\ee
where $F_{ab} = \partial_a v_b - \partial_b v_a$ with a vector field, 
\be
v_a = {1 \over 2} \epsilon^{ijk} v_a^{ijk}(\psi). 
\label{boson-v}
\ee
Note that the terms for $\partial^a v_a^{ljl}(\psi)$ in the variations (\ref{U1v-lambda(v)}), 
whose functional forms corresonds to the auxiliary fields $G = G(\psi)$ in Eq.(\ref{recombi-FG}), 
are absorbed into recombinations with respect to the auxiliary components $D^i$ as shown below. 

Under the conditions (\ref{U1condition}), the variations of $\tilde \lambda^i$ are expressed 
at least up to ${\cal O}(\psi^2)$ as 
\be
\delta_\zeta \tilde \lambda^i 
= \epsilon^{ijk} \left( \hat D^j - {i \over \alpha_{11}} \!\!\not\!\partial A^j \right) \zeta^k 
+ {i \over \alpha_{12}} \gamma_5 \!\!\not\!\partial \phi \zeta^i 
- {1 \over {2 \alpha_{13}}} \epsilon^{ab} F_{ab} \gamma_5 \zeta^i, 
\label{LSUSY-tlambda}
\ee
where bosonic components $\hat D^i$, $A^i$ and $\phi$ are defined by means of 
\be
\hat D^i = \left( D^i - {1 \over \alpha_{13}} \partial^a v_a^{jij} \right)(\psi), 
\ \ A^i = \left( {1 \over 2} M^{ijj} - M^{jij} \right)(\psi), 
\ \ \phi = {1 \over 2} \epsilon^{ijk} \phi^{ijk}(\psi). 
\label{boson-DAp}
\ee

The values of the constants $(a,b,c,d,e)$ in Eq.(\ref{tilde-lambda0}), which satisfy the relations (\ref{U1condition}), 
will be determined from considerations for LSUSY transformations of other component fields 
in the minimal $U(1)$ gauge supermultiplet. 
Moreover, as for higher-order terms of $\psi^i$ in the bosonic conponents (\ref{boson-v}) and (\ref{boson-DAp}) 
and the singlet spinor field in the helicity states (\ref{N3helicity}), whose leading term of $\psi^i$ 
is defined as 
\be
\chi = \xi^i \psi^i, 
\ee
it is expected that those higher-order terms are determined 
from general arguments as in Eqs. from (\ref{tilde-lambda0}) to (\ref{U1condition}), 
though most of the calculations are complicated. 

With the help of the study in the relation between the VA NLSUSY theory 
and the minimal $U(1)$ gauge supermultiplet for $d = 2$, $N = 3$ SUSY \cite{ST0}, 
we find that the following recombinations of $\psi^i$, 
\ba
\A \A 
\tilde D^i = \left\{ D^i - {1 \over \alpha_{13}} \partial^a v_a^{jij} 
- {1 \over {8 \alpha_{31}}} \Box (C^{ijjkk} - 4C^{jijkk}) \right\}(\psi) + {\cal O}(\psi^6), 
\label{t-D} \\
\A \A 
\tilde \lambda^i = \left[ \lambda^i - {i \over 2} \epsilon^{ijk} 
\left\{ {1 \over \alpha_{21}} \!\!\not\!\partial (\Lambda^{jkll} + \Lambda^{ljkl}) 
+ {i \over \alpha_{23}} \partial^a \Lambda_a^{ljkl} \right\} \right](\psi) 
+ {\cal O}(\psi^5), 
\label{t-lambda} \\
\A \A 
\tilde \chi = \left( \chi + {i \over {2 \alpha_{21}}} \!\!\not\!\partial \Lambda^{iijj} \right)(\psi) + {\cal O}(\psi^5), 
\label{t-chi} \\
\A \A 
\tilde A^i = \left( A^i - {\alpha_{11} \over {2 \alpha_{34}}} \partial^a C_a^{jijkk} \right)(\psi) + {\cal O}(\psi^6), 
\label{t-A} \\
\A \A 
\tilde \phi = \left( \phi 
+ {\alpha_{12} \over {4 \alpha_{34}}} \epsilon^{ab} \epsilon^{ijk} \partial_a C_b^{ijkll} \right)(\psi) + {\cal O}(\psi^6), 
\label{t-phi} \\
\A \A 
\tilde v^a = \left( v^a
- {\alpha_{13} \over {4 \alpha_{32}}} \epsilon^{ab} \epsilon^{ijk} \partial_b C_5^{ijkll} \right)(\psi) + {\cal O}(\psi^6). 
\label{t-v}
\ea
give LSUSY transformations in the minimal $U(1)$ gauge supermultiplet 
at least up to ${\cal O}(\psi^2)$ or ${\cal O}(\psi^3)$. 
Note that the recombinations (\ref{t-lambda}) are obtained from Eq.(\ref{tilde-lambda0}) 
by taking the values of the constants $(a,b,c,d,e)$ 
as $\displaystyle{\left( -{1 \over 2}, -{1 \over 2}, 0, {1 \over 2}, 0 \right)}$, 
%
which satisfy the conditions (\ref{U1condition}). 

Then, $U(1)$ gauge invariant and minimal expressions of LSUSY transformations 
for ($\tilde D$, $\tilde \lambda^i$, $\tilde \chi$, $\tilde A^i$, $\tilde \phi$, $\tilde v^a$) 
are confirmed at least up to ${\cal O}(\psi^2)$ or ${\cal O}(\psi^3)$ as follows; 
\ba
\A \A 
\delta_\zeta \tilde D^i = -i (\epsilon^{ijk} \bar\zeta^j \!\!\not\!\partial \tilde \lambda^k 
+ \bar\zeta^i \!\!\not\!\partial \tilde \chi), 
\nonu
\A \A 
\delta_\zeta \tilde \lambda^i 
= \epsilon^{ijk} \left( \tilde D^j - {i \over \alpha_{11}} \!\!\not\!\partial \tilde A^j \right) \zeta^k 
+ {i \over \alpha_{12}} \gamma_5 \!\!\not\!\partial \tilde \phi \zeta^i 
- {1 \over {2 \alpha_{13}}} \epsilon^{ab} \tilde F_{ab} \gamma_5 \zeta^i, 
\nonu
\A \A 
\delta_\zeta \tilde \chi 
= \left( \tilde D^i + {i \over \alpha_{11}} \!\!\not\!\partial \tilde A^i \right) \zeta^i, 
\nonu
\A \A 
\delta_\zeta \tilde A^i 
= \alpha_{11} (\epsilon^{ijk} \bar\zeta^j \tilde \lambda^k - \bar\zeta^i \tilde \chi), 
\nonu
\A \A 
\delta_\zeta \tilde \phi 
= -\alpha_{12} \bar\zeta^i \gamma_5 \tilde \lambda^i, 
\nonu
\A \A 
\delta_\zeta \tilde v^a 
= -\alpha_{13} (i \bar\zeta^i \gamma^a \tilde \lambda^i + \partial^a W_\zeta), 
\label{minLSUSY}
\ea
where a gauge transformation parametar $X$ in $\delta_\zeta \tilde v^a$ is 
\be
W_\zeta = {1 \over 2} \epsilon^{ijk} \left\{ {1 \over \alpha_{21}} \bar\zeta^i \Lambda^{jkll} 
- {1 \over 2} \left( {3 \over \alpha_{21}} \bar\zeta^l \Lambda^{ijkl} 
- {1 \over \alpha_{22}} \bar\zeta^l \gamma_5 \Lambda_5^{ijkl} 
+ {i \over \alpha_{23}} \bar\zeta^l \gamma^a \Lambda_a^{ijkl} \right) \right\}. 
\ee
We notice that the relations for the auxiliary fields $\Lambda^{ij}{}_A{}^k$ 
in Eqs. from (\ref{sub-Lambda2}) to (\ref{linear-combi}) are used 
in the derivation of the LSUSY transformations (\ref{minLSUSY}), 
in particular, for the bosonic components ($\tilde D$, $\tilde A^i$, $\tilde \phi$, $\tilde v^a$). 
Namely, those relations among the general auxiliary fields play the role 
not only in counting the d.o.f. for the component fields 
but also in transforming to the minimal $U(1)$ gauge supermultiplet from the general one. 

As is the case with the LSUSY transformations (\ref{N2U1v-v}) in the $N = 2$ $U(1)$ gauge supermultiplet, 
the $U(1)$ gauge transformation parameter $W_\zeta$ for $\tilde v^a$ in Eq.(\ref{minLSUSY}) 
leads to a relation (up to ${\cal O}(\psi^2)$), 
$\delta_{\zeta_1} W_{\zeta_2} - \delta_{\zeta_2} W_{\zeta_1} 
= - \theta(\zeta_1^i, \zeta_2^i; \tilde A^i, \tilde \phi, \tilde v_a)$, 
in the commutator algebra for  $\tilde v_a$. 
Thanks to this relation, a $U(1)$ gauge transformation term with the parameter $\theta$ 
is not induced in the commutation relation. 

The relation between a $N = 3$ LSUSY (free) action for the $U(1)$ gauge supermultiplet, which is denoted as $S_{N = 3\ {\rm gauge}}$, 
and the VA NLSUSY action (\ref{NLSUSYaction}) for $N = 3$ SUSY, i.e., 
\be
S_{N = 3\ {\rm gauge}}(\psi) + [\ {\rm a\ surface\ term}\ ] = S_{N = 3\ {\rm NLSUSY}} 
\label{N3U(1)-NLrelation}
\ee
has been shown in the Ref.\cite{ST0} at least up to ${\cal O}(\psi^3)$. 
In the relation (\ref{N3U(1)-NLrelation}), we define the LSUSY action $S_{N = 3\ {\rm gauge}}$ as 
\be
S_{N = 3\ {\rm gauge}} = \int d^2x \left\{ 
-{1 \over 4} (\tilde F_{ab})^2 + {i \over 2} \bar{\tilde \lambda}^i \!\!\not\!\partial \tilde \lambda^i 
+ {i \over 2} \bar{\tilde \chi} \!\!\not\!\partial \tilde \chi 
+ {1 \over 2} (\partial_a \tilde A^i)^2 + {1 \over 2} (\partial_a \tilde \phi)^2 
+ {1 \over 2} (\tilde D^i)^2 - {\xi^i \over \kappa} \tilde D^i \right\}, 
\label{N3U1action}
\ee
where $\tilde F_{ab} = \partial_a \tilde v_b - \partial_b \tilde v_a$ 
and the values of the constants ($\alpha_{11}$, $\alpha_{12}$, $\alpha_{13}$) are determined 
as $\alpha_{11}^2 = \alpha_{12}^2 = \alpha_{13}^2 = 1$ 
from the invariance of the action (\ref{N3U1action}) under the LSUSY transformations from (\ref{minLSUSY}). 
Note that the relative scales of the terms for the general auxiliary fields to the kinetic terms of the physical fields 
in the action (\ref{N3U1action}) are fixed by taking the values of 
($\alpha_{21}$, $\alpha_{23}$, $\alpha_{31}$, $\alpha_{32}$, $\alpha_{34}$, etc.) 
appropriately in the recombinations from (\ref{t-D}) to (\ref{t-v}).

\newpage

\section{Summary and discussions}

We have discussed the linearization of NLSUSY based on the commutator algebra (\ref{NLSUSYcomm}) 
and derived the general structure of the vector supermultiplets for extended LSUSY theories in $d = 2$. 
By defining the bosonic and fermionic component fields (\ref{b-comp}) and (\ref{f-comp}) 
from the set of the functionals (\ref{bosonic}) and (\ref{fermionic}), we have obtained 
the LSUSY transformations from (\ref{v-D}) to (\ref{v-C}) for vector supermultiplets with general auxilialy fields. 
Those LSUSY transformations are uniquely determined from the variations (\ref{variation1}) and (\ref{variation2}) 
under the commutation relation (\ref{NLSUSYcomm2}) as is the case with the linearization of NLSUSY in $d = 4$ \cite{MT1}. 
In the definition of the functionals (\ref{bosonic}) and (\ref{fermionic}), 
the fundamental determinant $\vert w \vert$ in the VA NLSUSY theory play the important role. 

The reduction from the general vector supermultiplets obtained in Section 3 to the $N = 2$ one gives some instructions 
for counting the d.o.f. of the bosonic and fermionic components up to the general auxiliary fields: 
Indeed, in Section 6, the relations from (\ref{sub-Lambda01}) to (\ref{sub-Lambda03}) and (\ref{sub-C}) 
lead the counting procedure (a), 
while the definition (\ref{N2-C}) of the auxiliary component $C$ gives its procedure (b). 
The minimal $U(1)$ gauge or scalar supermultiplet for $N = 2$ SUSY is obtained 
by means of appropriate recombinations of the functionals of the NG fermions as in Eqs.(\ref{recombi-lambda}) 
and (\ref{recombi-D}) or Eqs. from (\ref{recombi-chi}) to (\ref{recombi-AB}), 
which are similar to the case of the $d = 4$, $N = 1$ SUSY theory \cite{MT2} corresponding to the Wess-Zumino gauge. 

We have found in Section 6 that the d.o.f. of the general bosonic and fermionic components 
for the $N = 3$ vector supermultiplet are balanced as $44 = 44$ 
by using the procedure for counting the d.o.f. of the components, 
which is the simplest but a general example for the extended LSUSY theory with the general auxiliary-field structure. 
In particular, the relations (constraints) for each auxiliary fields ($\Lambda^{ij}{}_A{}^k$, \ $C^i{}_A{}^{jk}{}_B{}^l$, 
\ $\Psi^{ij}{}_A{}^{kl}{}_B{}^m$, \ $E^i{}_A{}^{jk}{}_B{}^{lm}{}_C{}^n$) are determined 
from their functional (composite) structure of the NG fermions. 

Moreover, by introducing the basic component fields (\ref{N3comp}) for the helicity states (\ref{N3helicity}), 
we have derived the $U(1)$ gauge supermultiplet for $N = 3$ SUSY in Section 7. 
The component fields ($\tilde D$, $\tilde \lambda^i$, $\tilde \chi$, $\tilde A^i$, $\tilde \phi$, $\tilde v^a$) 
in the $U(1)$ gauge supermultiplet are determined from the general recombinations of the functionals of the NG fermions 
in Eqs. from (\ref{t-D}) to (\ref{t-v}), 
which are a generalization of the Wess-Zumino gauge to the extended SUSY theories. 


The LSUSY transformations (\ref{minLSUSY}) for the $d = 2$, $N = 3$ minimal $U(1)$ gauge supermultiplet 
are obtained from the variations of the functionals from (\ref{t-D}) to (\ref{t-v}) 
under the NLSUSY transformations (\ref{NLSUSY}), where we have used the relations from (\ref{sub-Lambda2}) to (\ref{linear-combi}) 
for the auxiliary fields $\Lambda^{ij}{}_A{}^k$. 
That is, from the viewpoint of the commutator-based linearization of NLSUSY, 
the relations (constraints) for each auxiliary fields, which are determined from the behavior of the NG fermions, 
have the crucial role not only in counting the d.o.f. for the general component fields in LSUSY theories 
but also in transforming to the $U(1)$ gauge supermultiplet. 

\vspace{5mm}

\noindent
{\large\bf Acknowledgements} \\[2mm]
The author gratefully acknowledges Kazunari Shima for reading manuscript and valuable discussions and comments, 
e.g. in the relation of our work to the composite model interpretation in Refs.\cite{KS,KS0}.

\newpage

%
\newcommand{\NP}[1]{{\it Nucl.\ Phys.\ }{\bf #1}}
\newcommand{\PL}[1]{{\it Phys.\ Lett.\ }{\bf #1}}
\newcommand{\CMP}[1]{{\it Commun.\ Math.\ Phys.\ }{\bf #1}}
\newcommand{\MPL}[1]{{\it Mod.\ Phys.\ Lett.\ }{\bf #1}}
\newcommand{\IJMP}[1]{{\it Int.\ J. Mod.\ Phys.\ }{\bf #1}}
\newcommand{\PR}[1]{{\it Phys.\ Rev.\ }{\bf #1}}
\newcommand{\PRL}[1]{{\it Phys.\ Rev.\ Lett.\ }{\bf #1}}
\newcommand{\PTP}[1]{{\it Prog.\ Theor.\ Phys.\ }{\bf #1}}
\newcommand{\PTPS}[1]{{\it Prog.\ Theor.\ Phys.\ Suppl.\ }{\bf #1}}
\newcommand{\AP}[1]{{\it Ann.\ Phys.\ }{\bf #1}}


\begin{thebibliography}{100}

\bibitem{VA}
D.V. Volkov and V.P. Akulov, {\it Phys. Lett. B} {\bf 46} (1973) 109. 

\bibitem{WZ}
J. Wess and B. Zumino, {\it Phys. Lett. B} {\bf 49} (1974) 52. 

\bibitem{WB}
J. Wess and J. Bagger, {\it Supersymmetry and Supergravity 
(Second Edition, Revised and Expanded)} 
(Princeton University Press, Princeton, New Jersey, 1992). 

\bibitem{IK}
E.A. Ivanov and A.A. Kapustnikov, {\it J. Phys. A} {\bf 11} (1978) 2375. 

\bibitem{Ro}
M. Ro\v{c}ek, {\it Phys. Rev. Lett.} {\bf 41} (1978) 451. 

\bibitem{UZ}
T. Uematsu and C.K. Zachos, {\it Nucl. Phys. B} {\bf 201} (1982) 250. 

\bibitem{STT1}
K. Shima, Y. Tanii and M. Tsuda, {\it Phys. Lett. B} {\bf 525} (2002) 183. 

\bibitem{STT2}
K. Shima, Y. Tanii and M. Tsuda, {\it Phys. Lett. B} {\bf 546} (2002) 162. 

\bibitem{ST0}
K. Shima and M. Tsuda, {\it Phys. Lett. B} {\bf 641} (2006) 101. 

\bibitem{MT1}
M. Tsuda, {\it Phys. Lett. B} {\bf 764} (2017) 295. 

\bibitem{MT2}
M. Tsuda, {\it Mod. Phys. Lett. A} {\bf 31} (2016) 1650225. 

\bibitem{KS}
K. Shima, {\it Phys. Lett. B} {\bf 501} (2001) 237. 

\bibitem{ST1}
K. Shima and M. Tsuda, {\it Phys. Lett. B} {\bf 507} (2001) 260. 

\bibitem{KS0}
K. Shima, {\it Z. Phys. C} {\bf 18} (1983) 25; \\
K. Shima, {\it European Phys. J. C} {\bf 7} (1999) 341. 

\bibitem{ST2}
K. Shima and M. Tsuda, {\it Phys. Lett. B} {\bf 666} (2008) 410. 

\bibitem{ST3}
K. Shima and M. Tsuda, {\it Mod. Phys. Lett. A} {\bf 22} (2007) 1085 

\bibitem{ST4}
K. Shima and M. Tsuda, {\it Mod. Phys. Lett. A} {\bf 24} (2009) 185. 

\bibitem{ST5}
K. Shima and M. Tsuda, {\it Phys. Lett. B} {\bf 687} (2010) 89. 

\bibitem{STal}
K. Shima and M. Tsuda, {\it On the role of the commutator algebra for nonlinear supersymmetry}, 
arXiv:1607.05926[hep-th]. 

\bibitem{DVF}
P. Di Vecchia and S. Ferrara, {\it Nucl. Phys. B} {\bf 130} (1977) 93. 

\bibitem{ST6}
K. Shima and M. Tsuda, {\it Mod. Phys. Lett. A} {\bf 23} (2008) 1167. 

\bibitem{ST7}
K. Shima and M. Tsuda, {\it Mod. Phys. Lett. A} {\bf 23} (2008) 3149. 












%
%
%
%
%
%
%
%
%
%
%
%
%
%
%
%
%
%
%
%
%
%
%
%
%
%
%
%
%
%
%
%
%
%
%
%
%
%
%
%

\end{thebibliography}
\end{document}